\documentclass[12pt,preprint]{aastex}

\usepackage{amsmath}

\def\vec#1{\mbox{\boldmath $#1$}}
\usepackage{color}

\shorttitle{Data-Driven MHD Simulation of the Coronal Magnetic Field}
\shortauthors{Inoue $\&$ Hayashi}

\begin{document}


\title{An Evolution and Eruption of the Coronal Magnetic Field through a Data-Driven MHD Simulation}


\author{Satoshi Inoue}
\affil{Center for Solar-Terrestrial Research, New Jersey Institute of Technology, University Heights, Newark, NJ 07102-1982, USA}
\email{Satoshi.Inoue@njit.edu}
 
\author{Keiji Hayashi}
\affil{George Mason University, 4400 University Dr, Fairfax, VA 22030, USA}

\author{Takahiro Miyoshi}
\affil{Graduate School of Advanced Science and Engineering, Hiroshima University, 1-3-1 Kagamiyama, Higashihiroshima 739-8526, Japan}
       
    \begin{abstract}
    We present a newly developed data-driven magnetohydrodynamics (MHD) simulation code under a zero-$\beta$ approximation based on  a method proposed by Hayashi et al. 2018 
    and 2019.  Although many data-driven MHD simulations have been developed and conducted, there are not many studies on how accurately those simulations can reproduce the 
    phenomena observed in the solar corona. In this study, we investigated the performance of our data-driven simulation quantitatively using ground-truth data. The ground-truth data
    was produced by an MHD simulation in which the magnetic field  is twisted by the sunspot motions.  A magnetic flux rope (MFR) is created by the cancellation of the 
    magnetic flux at the polarity inversion line due to the converging flow on the sunspot, which eventually leads the eruption of the MFR.  We attempted to reproduce these dynamics 
    using the data-driven MHD simulation. The coronal magnetic fields are driven by the electric fields, which are obtained from a time-series of the photospheric magnetic field
    that is extracted from the ground-truth data, on the surface. As a result, the data-driven simulation could capture the subsequent MHD processes, the twisted coronal magnetic 
    field and formation of the MFR, and also its eruption. We report these results and compare with the ground-truth data, and discuss how to improve the accuracy and optimize 
    numerical method.
     \end{abstract}

    \section{Introduction}
    Active phenomena observed in the solar corona are widely believed to be driven by the coronal magnetic field (\citealt{2011LRSP....8....6S}). Thus, knowledge of how coronal magnetic 
    field evolves from the initial energy through its release is essential to understand these phenomena. The solar flares are one of the most common of these phenomena and are often 
    linked to the origin site of coronal mass ejections (CMEs).  CMEs bring lots of coronal gas and magnetic field from the Sun to interplanetary space and cause massive electromagnetic 
    disturbances in the Earth's magnetosphere when they collide. Therefore, how the coronal magnetic field evolves not only influences the magnetic environment of the solar corona, but also 
    extends this influence to near-Earth space and even to the furthest reaches of the heliosphere.
        
     Magnetic field observations have been conducted by ground and space-based observations which provide us a time-series of the photospheric magnetic field data in high 
     temporal and spatial resolutions. For instance, the solar dynamics observatory (SDO: \citealt{2012SoPh..275....3P}) provides the magnetic field every 12 minutes with 
     a spatial resolution of $0.5^{''}$, which is in Spaceweather HMI Active Patch format (\citealt{2014SoPh..289.3549B}). In the study of \cite{2017NatAs...1E..85W}, Goode Solar Telescope 
     (GST: \citealt{2012SPIE.8444E..03G}) at Big Bear Solar Observatory (BBSO) provides magnetic field at $0.08^{''}$ with 87 second cadence. However, the fact that magnetic field is only 
     measured at the photosphere is a technical problem, {\it i.e.}, three-dimensional (3D) magnetic field is not measured directly. A nonlinear force-free field extrapolation (NLFFF:{\it e.g.}, 
     \citealt{2016PEPS....3...19I}) is one of useful tool, which extrapolates a 3D magnetic filed numerically based on the photospheric magnetic field under a force-free approximation. 
     This technique has several achievements, for instance, time variation of the 3D magnetic structure before and after a flare (\citealt{2008ApJ...675.1637S}), a formation process of the 
     non-potential magnetic field leading to a solar flare ({\it e.g.}, \citealt{2011ApJ...738..161I},  \citealt{2012ApJ...748...77S}, \citealt{2014ApJ...780...55J}, \citealt{2020ApJ...890...84W}), and 
     analysis of solar flare onset in terms of time variations of  the 3D magnetic fields ({\it e.g.}  \citealt{2018ApJ...863..162M}, \citealt{2019ApJ...887..263K}, \citealt{2021ApJ...908..132Y}). 
     However, the NLFFF extrapolation only provides one snapshot of the magnetic field bounded by the force-free assumption and the essential problems still remain. For instance, the 
     photospheric magnetic field used as the boundary condition cannot satisfy the force-free state (\citealt{1995ApJ...439..474M}, \citealt{2020ApJ...898...32K}) and there is no guarantee of 
     a unique solution under the given boundary and initial conditions (\citealt{2020ApJ...895..105K}). 
     
    Although the NLFFF gives a snapshot of 3D magnetic field at specific time, a data-constrained magnetohydrodynamic (MHD) simulation, which uses the potential field, NLFFF and 
    non-force-free field as the initial conditions, can cover the evolution and dynamics of the magnetic field that is freed from an assumption of the equilibrium state ({\it e.g.}, 
    \citealt {2016PEPS....3...19I}, \citealt{2017ApJ...842...86M}, \citealt{2018NatCo...9..174I}, \citealt{2018ApJ...866...96J}, \citealt{2020ApJ...903..129P}, \citealt{2021PhPl...28b4502N}). 
    The data-constrained MHD simulation uses only one snapshot vector magnetic field.  According to \cite{2018NatCo...9..174I},  the normal component of the photospheric magnetic field is 
    fixed with time and the time-dependent horizontal magnetic fields are derived from the induction equation, so that the evolution of the tangential components is not exactly consistent with 
    the time-series of the photospheric magnetic field.   Nevertheless, these simulations reproduce the observations quantitatively, such as distribution of the flare ribbons and 
    erupting direction of the magnetic flux rope ({\it e.g.,} \citealt{2014ApJ...788..182I},  \citealt{2017ApJ...850....8J}, \citealt{2017ApJ...842...86M}).   
             
     A data-driven MHD simulation drives the coronal magnetic field where a time-series of the photospheric magnetic field is applied to the boundary condition (\citealt{2020ApJ...890..103T},  
     \citealt{2022Innov...300236J}). This method is more advanced than the data-constrained MHD simulation. The problem is how to treat the boundary condition, {\it i.e.,} how to treat the 
     induction equation at bottom boundary. Note that we consider very simple situation under zero-$\beta$ approximation without taking into account the time-varying density 
     and pressure at the bottom surface.  The induction equation is written as follows in an ideal MHD framework, 
    \begin{equation}
    \frac{\partial \vec{B}}{\partial t} = - \vec{\nabla}\times\vec{E} = \vec{\nabla}\times(\vec{v}\times\vec{B}),
    \label{dd_in_eq}
    \end{equation}
    where $\vec{B}$, $\vec{E}$ and $\vec{v}$ are magnetic field, electric field, and plasma velocity, respectively. From Eq.(\ref{dd_in_eq}), we have several ways to drive the coronal 
    magnetic field above, for instance, the electric field is given at the bottom boundary (\citealt{2010ApJ...715..242F}, \citealt{2020ApJS..248....2F}, \citealt{2014ApJ...795...17K},  
    \citealt{2017SoPh..292..191L}, \citealt{2018ApJ...855...11H}). $\vec{E}$ has gauge invariance {\it i.e.}, it should be described as $\vec{E}=\vec{E}_{I} + \vec{\nabla}\phi$ which 
    implies that the given electric field and the evolution of the coronal magnetic field depend on the gauge invariance. Nevertheless, \cite{2012ApJ...757..147C}, 
    \cite{2019SoPh..294...41P} and \cite{2020A&A...644A..28P}  reproduced various observational features by choosing the gauge with ingenuity. For example,  \cite{2019SoPh..294...41P} 
    compares the magnetic field liens with the Extreme Ultraviolet (EUV) images and  \cite{2020A&A...644A..28P} superimposes the field lines on the images for a more rigorous comparison. 
    The structure of the magnetic field lines is in good agreement with the structure of the magnetic field lines inferred from the images. Furthermore,  \cite{2012ApJ...757..147C} 
    reproduced the back reaction of the eruption that enhances the photospheric magnetic field after the eruption. Recently, \cite{2021ApJ...909..155K} derived the photospheric velocity by 
    inversely solving the induction equation by imposing physical constraints on the gauge and driving the coronal magnetic field using the obtained velocity. Several methods directly give 
    either the magnetic field or velocity, or both in their data-driven simulation despite both the magnetic field and the velocity are given at the boundary conditions is an over-specification for 
    an MHD.  For example, the velocity is derived from the algorithm of Differential Affine Velocity Estimator for Vector Magnetogram (DAVE4VM:\citealt{2008ApJ...683.1134S}). Although the 
    detailed method is different for each, they have produced a non-potential magnetic field prior to the flare or the eruption of the MFR well (\citealt{2016NatCo...711522J}, 
    \citealt{2017ApJ...838..113L}, \citealt{2019A&A...626A..91L}, \citealt{2019ApJ...870L..21G},  \citealt{2019ApJ...871L..28H}, \citealt{2021FrP.....9..224J}).  
    
    Recently, \cite{2018ApJ...855...11H} and \cite{2019ApJ...871L..28H} proposed a new method for the data-driven MHD simulation in which three electric fields are defined at the bottom 
    boundary, and a half-grid above and below that completely reproduce the photospheric magnetic field through the induction equation. \cite{2019ApJ...871L..28H} includes a 
    data-driven simulation 
    in which the coronal magnetic field  is driven by the velocity. Following their method, the velocity, which is obtained from DAVE4VM, is given on the bottom surface while the magnetic 
    field is driven by the velocity through the MHD process. Their simulation was applied to solar active region (AR) 11158 and reproduced the evolution  toward the non-potential magnetic 
    field starting with the potential field. Eventually the magnetic fields prior to the flare were obtained, which were similar to coronal magnetic fields inferred from the 
    EUV observations. In this study we develop a 
    data-driven MHD code based on the method proposed by \cite{2018ApJ...855...11H} . The code is designed with a zero-$\beta$ approximation because the magnetic pressure is dominant 
    over the gas pressure in the lower solar corona, and constructed using a finite-differential method based on \citealt{2014ApJ...788..182I}. The purpose of this study is to test the accuracy 
    of this data-driven MHD simulation to compare with the ground-truth data including evolution of  the coronal magnetic fields during the energy buildup and eruption process. 
    \cite{2017ApJ...838..113L}  and \cite{2020ApJ...890..103T} discussed accuracy of the data-driven simulations by comparison to the ground-truth data. \cite{2017ApJ...838..113L} 
    examined the sensitivity of data-driven simulation results to the cadence of the input boundary driving-data maps. In their study, all MHD variables are given at the bottom-boundary.
    They found that the cadence is indeed important for reproducibility. \cite{2020ApJ...890..103T} compares the simulation results from the participating data-driven simulation models.
    In their comparison, the models are given only the information on the boundary magnetic field at a low cadence and yield very different solutions from the ground-truth solution.
    Our electric field driven model uses only the information on the solar-surface magnetic field. Therefore, our newly developed code also needs the accuracy test.
    
      From the two aforementioned works, in this present study, the target ground-truth three-dimensional solution and the bottom-boundary time-dependent magnetic field data maps would 
    be prepared with the same MHD code but driven with variables other than magnetic field. We followed the velocity-driven MHD model procedures by  \cite{2000ApJ...529L..49A} and 
    \cite{2003ApJ...585.1073A}  for generating the ground-truth solutions and the time-dependent magnetic-field data maps. This ground-truth data includes the eruption of the magnetic field 
    which is different from previous works by \cite{2017ApJ...838..113L}  and \cite{2020ApJ...890..103T}. 
    
     Note that the electric field obtained from the inversion of the
    induction equation is not uniquely determined due to the uncertainty of the gauge, hence it is not guaranteed that the electric field will indeed reproduce the evolution of the magnetic field 
    in the ground-truth data produced by velocity-driven MHD. In addition, the electric-field inversion is designed to utilize only (temporal difference of) the magnetic field data on the bottom 
    boundary surface at two consecutive sampling instants. In other word, the method in \cite{2018ApJ...855...11H}  is not designed to accommodate information on the temporal evolution of 
    the ground-truth magnetic field or the three-dimensional structure in the whole volume. Given these caveats above, it is crucially important to assess how well the electric field-driven MHD 
    models using the electric-field inversion can yield the ground-truth solution. In this work, we quantitatively evaluate the results of the data-driven MHD simulation by comparing with the 
    ground-truth data.  We further discuss the numerical treatment of implementation method, for which intrinsically we are allowed to make somewhat arbitrary choice in, for example, spatial 
    differencing.
                   
    The rest of this paper is constructed as follows. The numerical method is described in section 2. The results and discussion are presented in section 3 and section 4. 
    Finally, our conclusion are summarized in section 5. 

    \section{Numerical Method}  
    \subsection{Basic Equations}
    We solve the following zero-$\beta$ MHD equation to make the ground-truth data and also conduct the data-driven simulation,
        
  \begin{equation}
  \frac{\partial \rho}{\partial t} = -\vec{\nabla}\cdot(\rho \vec{v})+\zeta\vec{\nabla}^2\rho,
  \label{mass_eq}
  \end{equation}

  \begin{equation}
  \frac{\partial \vec{v}}{\partial t} 
                        = - (\vec{v}\cdot\vec{\nabla})\vec{v}
                          + \frac{1}{\rho} \vec{J}\times\vec{B}
                          + \nu\vec{\nabla}^{2}\vec{v},
  \end{equation}

  \begin{equation}
  \frac{\partial \vec{B}}{\partial t} 
                        =  \vec{\nabla}\times(\vec{v}\times\vec{B})
                        +  \eta \vec{\nabla}^2\vec{B}
                        -  \vec{\nabla}\phi, 
  \label{induc_eq}
  \end{equation}

  \begin{equation}
  \vec{J} = \vec{\nabla}\times\vec{B},
  \end{equation}
  
  \begin{equation}
  \frac{\partial \phi}{\partial t} + c^2_{h}\vec{\nabla}\cdot\vec{B} 
    = -\frac{c^2_{h}}{c^2_{p}}\phi,
  \label{div_eq}
  \end{equation}
  where $\vec{B}$ is the magnetic flux density, $\vec{v}$ is the velocity, $\vec{J}$ is the electric current density, $\rho$ is the pseudo density, and $\phi$ is the convenient potential 
  to remove errors derived from $\vec{\nabla}\cdot \vec{B}$ (\citealt{2002JCoPh.175..645D}), respectively. $\rho_0$ corresponds to the initial density. The length, magnetic field, 
  density, velocity, time, and electric current  density are normalized by   $L^{*}$ , $B^{*}$, $\rho^{*}$ , $V_{\rm A}^{*}\equiv B^{*}/(\mu_{0}\rho^{*})^{1/2}$, where $\mu_0$ is the 
  magnetic permeability, $\tau_{\rm A}^{*}\equiv L^{*}/V_{\rm A}^{*}$, and $J^{*}=B^{*}/\mu_{0} L^{*}$.  $\nu$ and $\eta$ are viscosity and resistivity, fixed by $1.0 \times 10^{-3}$ 
  and $1.0 \times 10^{-5}$, respectively,  and the coefficients $c_h^2$, $c_p^2$ in Eq.(\ref{div_eq}) also fix the constant value, 0.04 and 0.1, respectively. $\zeta$ is a diffusion  
  coefficient of the density which avoids the sudden variation of the density to  improve the robustness of the simulation, where $\zeta$ is set to $1.0 \times 10^{-4}$ in this study. 
  A numerical box of 1.0 $\times$ 1.0 $\times$ 1.0, which is given in its non-dimensional value, is divided by 320 $\times$ 320 $\times$ 320 grid points. 

    \subsection{Ground-truth data} 
\begin{figure}
  \epsscale{1.}
  \plotone{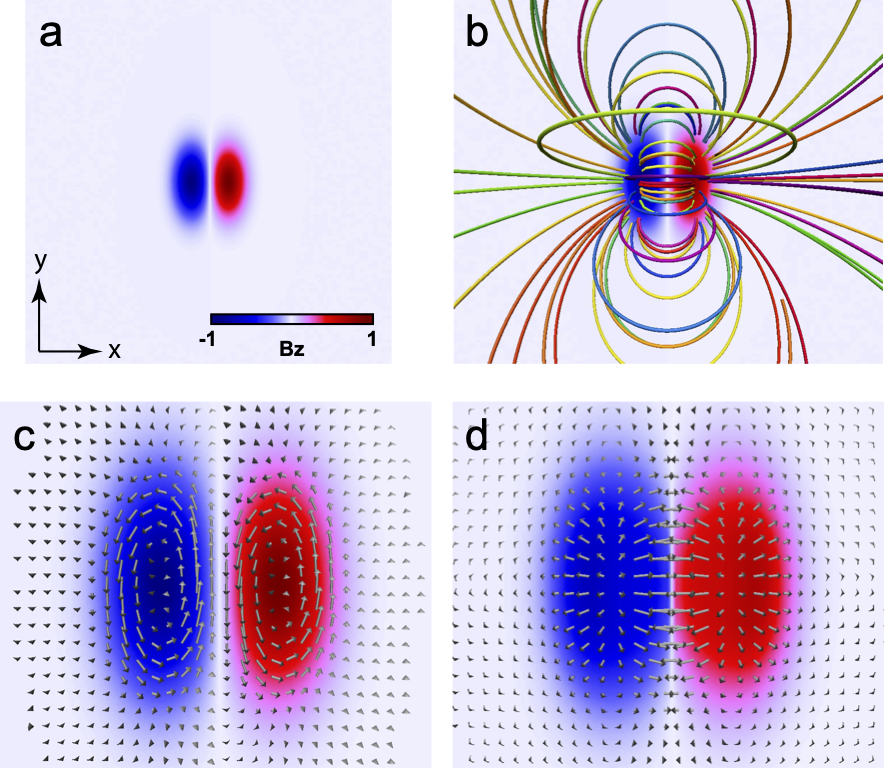}
  \caption{
          {\bf a} A bipolar magnetic field given at the bottom surface where the red and blue correspond to the positive and negative 
                     polarities, respectively. The full simulation region is plotted in X- and Y- directions.
          {\bf b} The potential magnetic field which is extrapolated from the bipolar magnetic field is shown in {\bf a}. This is used as the 
                    initial condition of the MHD simulation.
          {\bf c}  Twisting motion of the velocity given on the bipolar field, which can keep the distribution of the bipolar field. 
          {\bf d}  Diverging motion of the velocity given on the bipolar field. This velocity plays a role as converging motion at the PIL.   }
  \label{f1}
  \end{figure}

    We first make ground-truth data according to an MHD simulation done by \cite{2000ApJ...529L..49A} and \cite{2003ApJ...585.1073A} to verify the reproducibility of the energy storage 
    and release processes of the coronal magnetic field by our data-driven MHD simulation. We first set a simple bipole magnetic field like sunspots as shown in Fig.\ref{f1}a, from which 
    the potential field is extrapolated in the 3D space following the Green function method (\citealt{1982SoPh...76..301S}) as shown in Fig.\ref{f1}b. The initial velocity and density are set 
    as $|\vec{v}|=0.0$ and $\rho_0=1.0$ in the whole space, respectively.

    Next we impose a twisting motion on the photosphere according to, 
    \begin{equation}
    \vec{v}_h^{(b)}|_{z=0} = (v_x, v_y) =\left(-\frac{\partial \psi}{\partial x}, \frac{\partial \psi}{\partial y}\right),
    \end{equation}    
    where $\psi$ is a steam function which satisfies $ \vec{v}_h^{(b)}|_{z=0} = \vec{z}\times\vec{\nabla}_h\psi$. We give $\psi$ the following formula, 
    \begin{displaymath}
    \psi(x,y,t) = \gamma(t)\{B^{(b)}_z\}^2e^{\left\{\frac{-\left(B_z^{(b)}\right)^2 - \left(B^{(b)}_{z:max}\right)^2}{\left(B^{(b)}_{z:max}\right)^2}\right\}},
    \end{displaymath}
    and 
    \begin{displaymath}
    \gamma(t)=-0.5\tanh\left\{2.0\frac{\left(t-t_{cri}\right)-1.0}{0.5}\right\} + 0.5, 
    \end{displaymath}
    where $B_z^{(b)}$ and $B_{z:max}^{(b)}$ correspond to the magnetic field measured at the photosphere and the maximum value, respectively. If time $t$ is beyond the critical time 
    set at $t_{cri}$ where we set $t_{cri}=9.0$, $\gamma(t)$ quickly falls to zero.  The twisting motion is shown in Fig.\ref{f1}c. It is convenient to give $\psi$ as a function of $B_z$ 
    because the twisting motion is imposed along the contour of $B_z$, {\it i.e.}, $(\vec{v}_h\cdot\vec{\nabla})B_z=0$ is promised regardless of $B_z$. Thus the distribution of $B_z$ 
    will not be changed by the twisting motion. Since we assume $v_z$=0 at the bottom surface as well as $(\vec{v}_h\cdot\vec{\nabla})B_z=0$, these conditions satisfy 
    $\partial_t B_z = 0$. Therefore, normal flux is not transported across the bottom surface during the evolution and our simulation can keep $\int {\bf B}\cdot d{\bf S}=0$. 
    When $\gamma(t)$ falls to zero, the twisting motion  at the photosphere comes to a complete halt. Afterward, the magnetic field is relaxed for a while, {\it.i.e.}, no external motion is 
    imposed. The horizontal magnetic components $B_x$ and $B_y$ at the photosphere are following the equations during the twisting motion and relaxation process after the twisting 
    motion is over,
    \begin{displaymath}
   \frac{\partial B^{(b)}_x}{\partial t}=   - \left(\frac{\partial E_z}{\partial y} - \frac{\partial E_y}{\partial z}\right) - \frac {\partial \phi}{\partial x}, 
   \end{displaymath}
   \begin{equation}
   \frac{\partial B_y^{(b)}}{\partial t} =  - \left(\frac{\partial E_x}{\partial z} - \frac{\partial E_z}{\partial x} \right) - \frac{\partial \phi}{\partial y}, 
   \label{mag_bc_y_0}
   \end{equation}
    while $B_z$ is fixed. $E_x$ and $E_y$ represent the electric field in $x$ and $y$ components which are described as $(-\vec{v}\times\vec{B})\cdot \vec{x}$ and 
    $(-\vec{v}\times\vec{B})\cdot {\vec y}$, respectively. The $v_z$ and density $\rho$ are fixed with initial value at the bottom boundary and $\phi$ is imposed on the Neumann condition 
    throughout the simulation. Note that, during the relaxation process, $E_z$ at the bottom surface is zero, {\it i.e.}, $\partial_x E_z$ = $\partial_y E_y$=0 because the evolutions of 
    $v_x$ and $v_y$ are halted there ($v_x$ = $v_y$ = 0).
      
    Finally, after the relaxation, we impose a diverging motion on the sunspot, which corresponds to a converging motion around the PIL, as shown in Fig.\ref{f1}d. The diverging 
    motion is given as the following formula ({\it e.g.}, \citealt{2014ApJ...780..130X}), 
     \begin{displaymath}
    v_x^{(b)} = - \frac{\partial |B^{(b)}_z|}{\partial x}e^{\left\{\frac{-(x-0.5L_x)^2}{x_d^2}\right\}},
    \end{displaymath}
    \begin{equation}
    v_y^{(b)} = - \frac{\partial |B^{(b)}_z|}{\partial y}e^{\left\{\frac{-(y-0.5L_y)^2}{y_d^2}\right\}},
    \end{equation}    
    where $x_d$=$y_d$ =0.1 is given. The magnetic field at the bottom boundary follows an induction equation in each component as below, 
    \begin{displaymath}
   \frac{\partial B^{(b)}_x}{\partial t}=   - \left(\frac{\partial E_z}{\partial y} - \frac{\partial E_y}{\partial z}\right) + \eta\frac{\partial^2 B^{(b)}_x}{\partial x^2} - \frac {\partial \phi}{\partial x}, 
   \end{displaymath}
   \begin{displaymath}
   \frac{\partial B_y^{(b)}}{\partial t} =  - \left(\frac{\partial E_x}{\partial z} - \frac{\partial E_z}{\partial x}\right)   + \eta\frac{\partial^2 B^{(b)}_y}{\partial x^2} - \frac{\partial \phi}{\partial y}, 
   \end{displaymath}
    \begin{equation}
    \frac{\partial B_z^{(b)}}{\partial t} =  -\left(\frac{\partial E_y}{\partial x} - \frac{\partial E_x}{\partial  y}\right) +  \eta\frac{\partial^2 B^{(b)}_z}{\partial x^2} - \frac{\partial \phi}{\partial z}, 
    \label{mag_bc_y}
    \end{equation}
    note that the diffusion is given in the direction of converging motion (in this study, it corresponds to $x$ direction) to make the simulation more robust. This diverging motion is 
    continuously imposed on the sunspot until the end of the simulation. In both cases, twisting motion and diverging motion, we normalized the velocity $\vec{v}^{(b)}$ by the 
    maximum value $|v_{y:max}^{(b)}|$  and multiplied by 0.01, {\it i.e.}, $\vec{v}^{(b)} \Rightarrow 0.01\vec{v}^{(b)}/|v_{y:max}^{(b)}|$, before being used in Equations (\ref{mag_bc_y_0}) 
    and (\ref{mag_bc_y}), so that the maximum absolute value of the velocity is set as 0.01 at the bottom surface.

    \subsection{Data-driven Simulation}
    \subsubsection{Three Electric Fields Driving the Coronal Magnetic Field}
    In this study, a data-driven MHD simulation is performed according the method proposed by \cite{2018ApJ...855...11H} and \cite{2019ApJ...871L..28H}. We will now briefly describe  
     this method. In this method the coronal magnetic field is driven by electric fields given at the bottom.   In order to determine the electric fields that derive the time 
    evolving observed  photospheric magnetic field through an induction equation, three types of electric fields are assumed to satisfy the following induction equations, 
   \begin{equation}
    \frac{\partial B_z}{\partial t}  = - \vec{z}\cdot\vec{\nabla}\times\vec{E}^{(1)}, 
    \label{in_eq1}
    \end{equation}

    \begin{equation}
    \frac{\partial \vec{B}_{h:df}}{\partial t}  = - \vec{\nabla}\times\vec{E}^{(2)},
    \label{in_eq2}
    \end{equation}

   \begin{equation}
   \frac{\partial \vec{B}_{h:cf}}{\partial t}  = - \vec{\nabla}\times\vec{E}^{(3)},
   \label{in_eq3}
   \end{equation}
   where the horizontal field ($\vec{B}_h$) at the photosphere is decomposed into $\vec{B}_{h:df}$  and $\vec{B}_{h:cf}$, {\it i.e.}, $\vec{B}_h =\vec{B}_{h:df} + \vec{B}_{h:cf}$ 
   where these satisfy $\vec{\nabla}_h\cdot\vec{B}_{h:df}=0$ and $\vec{\nabla}_h\times\vec{B}_{h:cf}=0$, respectively.

    Following  \cite{2018ApJ...855...11H}, three Poisson equations are derived from the above three induction equations. First, when $\vec{z}\cdot\vec{\nabla}\times$ is multiplied by the 
    first induction equation (\ref{in_eq1}), the first Poisson equation is obtained as follows, 
    \begin{equation}
    \vec{\nabla}^2 \Phi^{(1)} = - \frac{\partial B_z}{\partial t}, 
    \label{poisson_1}
    \end{equation}
    where  we make the following assumption, $\vec{E}^{(1)} = \vec{z} \times \vec{\nabla}\Phi^{(1)}$, {\it i.e.}, $\partial E_z^{(1)}/\partial z = 0$. Since the right hand side of 
    Eq.(\ref{poisson_1}) can be calculated from the photospheric magnetic field, we can solve the Eq. (\ref{poisson_1}) under an appropriate boundary condition and 
    obtain $\vec{E}^{(1)}$ to determine $B_z$ at the photosphere.

    The second Poisson equation is 
    \begin{equation} 
    \begin{split}
    \vec{\nabla}^2\Phi^{(2)} &  = -\vec{z}\cdot\left(\vec{\nabla}\times\frac{\partial\vec{B}_{h:df}}{\partial t}\right)  \\
                                          &  = -\vec{z}\cdot\left(\vec{\nabla}\times\frac{\partial\vec{B}_h}{\partial t}\right), 
    \end{split}
    \label{poisson_2}
    \end{equation}
    where $ \vec{E}^{(2)} = -\vec{z}\Phi^{(2)}$ is assumed. The advantage of this assumption is that it does not change the normal component of the magnetic field at the photosphere because it is 
    determined by $E_x^{(1)}$ and $E_y^{(1)}$. Namely, if we further give $E_x^{(2)}$ and $E_y^{(2)}$, $B_z$ changes and deviates from the observation. This assumption is able to avoid 
    the conflict of $B_z$ being obtained from $\vec{E}^{(1)}$ and $\vec{E}^{(2)}$. The detailed derivation process of Eq.(\ref{poisson_2}) is described in Appendix (\ref{Apex_PE2}). 
 
    The third Poisson equation is 
    \begin{equation}
    \begin{split}
    \vec{\nabla}_h^2 \Phi^{(3)} & = \frac{1}{2}\Delta z \vec{\nabla}_h\cdot\left(\frac{\partial \vec{B}_{h:cf}}{\partial t}\right) \\
                                              &  = \frac{1}{2}\Delta z \vec{\nabla}_h\cdot\left(\frac{\partial \vec{B}_{h}}{\partial t}\right),
    \label{poisson_3}
    \end{split}
    \end{equation} 
    where $\Delta z$ is the grid interval and we make the following assumption, $\vec{E}^{(3)}|_{z=+\frac{1}{2}\Delta z} = \vec{z} \times \vec{\nabla}_h\Phi^{(3)}$.  Note that  since $B_z$ is perfectly 
    derived from $E_x^{(1)}$ and $E_y^{(1)}$, $\vec{E}^{(3)}$ is an over condition to determine $B_z$.  Therefore, $\vec{E}^{(3)}$ is defined at locations half-grid above and below the photosphere 
    as follows, 
    \begin{equation}
    \vec{E}^{(3)}|_{z=+\Delta z/2} = - \vec{E}^{(3)}|_{z=-\Delta z/2}.
    \label{eq_e3_1}
    \end{equation}
    $\Phi^{(3)}$ is defined at the plane half-grid above the bottom surface but it is determined by given horizontal magnetic field at the bottom surface, so $\Delta z$ plays the role of the bridge 
    between the left and right handed values defined at each different height. The detailed derivation process of Eq.(\ref{poisson_3}) is described in Appendix (\ref{Apex_PE3}).
    Eventually,  we obtain $\vec{E}^{(1)}$, $\vec{E}^{(2)}$ and $\vec{E}^{(3)}$ through the three Poisson equations (\ref{poisson_1}), (\ref{poisson_2}) and 
     (\ref{poisson_3}). The positional relationship of the three electric fields is summarized in Fig.\ref{f2}.

     \subsubsection{How to drive the magnetic field at the bottom surface}
    The magnetic field at the bottom surface is driven by $\vec{E}^{(1)}=(E_x^{(1)}, E_y^{(1)}, 0)$, $\vec{E}^{(2)}=(0, 0, E_z^{(2)})$ and $\vec{E}^{(3)}=(E_x^{(3)}, E_y^{(3)}, 0)$
    where $\vec{E}^{(1)}$ and $\vec{E}^{(2)}$  are defined at the bottom surface while $\vec{E}^{(3)}$ is defined one half-grid above (and also below) the bottom surface. Note that, following 
    \cite{2018ApJ...855...11H}, $\vec{E}^{(1)}$ is also placed at the surface one half-grid above (and also below) the bottom, taking into account conservation of the normal flux. The 
    different from $\vec{E}^{(3)}$ is that we assume $\vec{E}^{(1)}_{z=+ \Delta z/2}$=$\vec{E}^{(1)}_{z=-\Delta z/2}$, so its z derivative becomes zero.  This positional relationship is 
    represented in Fig.\ref{f2}. The time derivative of $B_z$ is written as 
    \begin{equation}
    \frac{\partial B_z}{\partial t} = -\left(\frac{\partial E_y^{(1)}}{\partial x}-\frac{\partial E_x^{(1)}}{\partial y}\right),
    \end{equation} 
     and since $\vec{E}^{(2)}$ only has a z-component, $\vec{B}_{h:df}$ is described as 
    \begin{equation}
    \frac{\partial \vec{B}_{h:df}}{\partial t} = \left(-\frac{\partial E_z^{(2)}}{\partial y}, \frac{\partial E_z^{(2)}}{\partial x}\right),
    \end{equation}
    where all the components are defined in $x$ and $y$ components, thus we omit the $z$ component. Regarding the time derivative of $\vec{B}_{h:cf}$,  by taking into account the 
    assumption of $\vec{E}^{(3)}$, $E_z^{(3)}$ equals to zero and $\vec{E}^{(3)}$ is defined as the plane half-grid above and below the bottom surface. The x-component of 
    $\partial_t \vec{B}_{h:cf}^{(3)}$ is written as
    \begin{displaymath}
    \begin{split}
    \left(\frac{\partial B_{x:cf}^{(3)}}{\partial t}\right)_{z=0} & =  \frac{\partial E_y^{(3)}}{\partial z} \\
                                                                                   & = \frac{E_y^{(3)}|_{+\Delta z/2} - E_y^{(3)}|_{-\Delta z/2}}{\Delta z} \\
                                                                                   & = \frac{2}{\Delta z}E_y^{(3)},
    \end{split} 
    \end{displaymath}
    similarly, 
    \begin{displaymath}
    \left(\frac{\partial B_{y:cf}^{(3)}}{\partial t}\right)_{z=0}   =  - \frac{2}{\Delta z}E_x^{(3)},     
    \end{displaymath}    
    where the details are described in the Appendix of \cite{2018ApJ...855...11H}. Therefore, 
    \begin{displaymath}
    \frac{\partial B_x}{\partial t} = -\left(\frac{\partial E_z^{(2)}}{\partial y} -\frac{2}{\Delta z}E_y^{(3)}\right),
    \end{displaymath}
    \begin{equation}
     \frac{\partial B_y}{\partial t} = -\left(-\frac{\partial E_z^{(2)}}{\partial x} + \frac{2}{\Delta z}E_x^{(3)}\right),
     \label{eq_bottom}
     \end{equation}
     \begin{displaymath}
      \frac{\partial B_z}{\partial t} = -\left(\frac{\partial E_y^{(1)}}{\partial x}-\frac{\partial E_x^{(1)}}{\partial y}\right),
     \end{displaymath}
     are given at the bottom surface.
    
     \subsection{Numerical method of the data-driven simulation}
\begin{figure}
  \epsscale{1.}
  \plotone{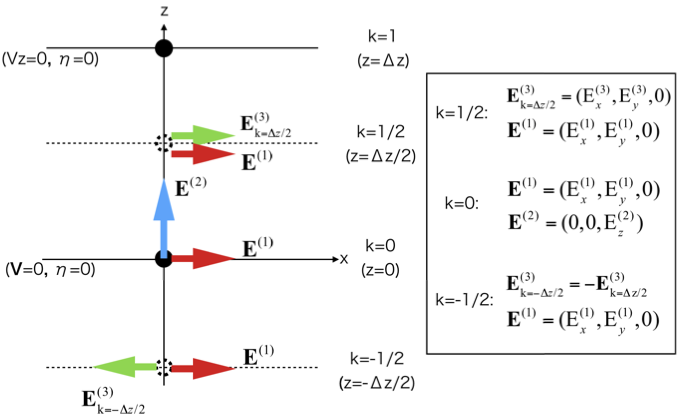}
  \caption{ 
                The positions of each electric field, $\vec{E}^{(1)}$, $\vec{E}^{(2)}$ and $\vec{E}^{(3)}$ that drive the coronal magnetic field, 
                proposed by \cite{2018ApJ...855...11H}. $z=$0 corresponds to the bottom boundary (photosphere) and $z=\Delta z$ corresponds to 
                a location one grid above the bottom surface where $\Delta z$ is a distance between neighboring grids set in this study.  All physical 
                values are defined at the black circles in this simulation where $k$ takes integer value. $\vec{E}^{1}$ and $\vec{E}^{3}$  are 
                given at $k=\pm{1/2}$. The right square summarizes the electric fields which are given at $k=1$ and $k=\pm{1/2}$.
          }
  \label{f2}
  \end{figure}

    The same MHD equations (Eq.(\ref{mass_eq})-Eq.(\ref{div_eq})) are applied to the data-driven MHD simulation while the electric fields, $\vec{E}^{(1)-(3)}$  are given at the bottom 
    and a half-grid above and  below as shown in Fig.\ref{f2}.  Since the location that is one grid above the bottom ($k$=1 shown in Fig.\ref{f2}) 
    corresponds to the practical bottom boundary, the derivation of the magnetic field at this position ($k$=1) is important for the data-driven simulation in this study. Especially, the 
    derivative in z direction introduced in the induction equation should be handled with care because it includes the difference between non-MHD components ($\vec{E}^{(1)}$ and 
    $\vec{E}^{(3)}$) which are placed at $k$=0 or 1/2  and the electric fields obtained from MHD process ($-\vec{v}\times\vec{B}$) which are placed at $k$=2.        
    
    According to a second order finite differential method, the derivative of E in the z-direction at $k$=1 in the induction equation is written as, 
    \begin{displaymath}
    \begin{split}
    \frac{\partial E}{\partial z}|_{k=1} = & \frac{E_{k=2}-E_{k=0}}{2\Delta z}, \\
                                                      = & \frac{E_{k=2} - E_{k=1}}{2\Delta z} + \frac{E_{k=1} - E_{k=0}}{2\Delta z}, \\
                                                      = & \frac{E_{k=2} - E_{k=1}}{2\Delta z} + \left(\frac{E_{k=1} + E_{k=0}}{2\Delta z}\right) - \frac{2E_{k=0}}{2\Delta z}, \\
                                                      = & \frac{E_{k=2} - E_{k=1}}{2\Delta z} + \left(\frac{E_{k=\frac{1}{2}}}{\Delta z} - \frac{E_{k=0}}{\Delta z}\right), 
    \end{split}
    \end{displaymath}
    where $E$ is either the $x$ or $y$ component and we use the following approximation, 
    \begin{equation}
    E_{k=\frac{1}{2}}=\frac{E_{k=1} + E_{k=0}}{2\Delta z},
    \label{OH_E}
    \end{equation}
    to obtain the last equation. When we take into account $E_{k=\frac{1}{2}}= E^{(1)}+E^{(3)}$ and $E_{k=0} = E^{(1)}$,  we obtain
    \begin{equation}
    \frac{\partial E}{\partial z}|_{k=1} = \frac{E_{k=2} - E_{k=1}}{2\Delta z}  + \frac{E^{(3)}}{\Delta z}. 
    \label{type-a}
    \end{equation} 
    
    On the other hand, we can write down the different formula as follows, 
    \begin{displaymath}
    \begin{split}
    \frac{\partial E}{\partial z}|_{k=1} = & \frac{E_{k=2}-E_{k=0}}{2\Delta z}, \\
                                                      = & \frac{E_{k=2} + E_{k=1}}{2\Delta z} - \frac{E_{k=1} + E_{k=0}}{2\Delta z}, \\    
                                                      = & \frac{E_{k=2} + E_{k=1}}{2\Delta z}  - \frac{E_{k=\frac{1}{2}}}{\Delta z},                                                                
    \end{split}
    \end{displaymath}
    where $E_{k=\frac{1}{2}}$ is approximated in Eq.(\ref{OH_E}). Eventually, we obtain the following equation,
    \begin{equation}
    \frac{\partial E}{\partial z}|_{k=1} = \frac{E_{k=2} + E_{k=1}}{2\Delta z}  - \frac{E^{(1)}+E^{(3)}}{\Delta z}.
    \label{type-b}
    \end{equation}    
    
    Hereafter, the former written in Eq.(\ref{type-a}) is denoted as Type-A and the later (Eq.(\ref{type-b})) is denoted as Type-B. We will mainly show the results obtained from 
    Type-A and discuss the difference in results between Type-A and Type-B. 
        
\begin{figure}
  \epsscale{1.}
  \plotone{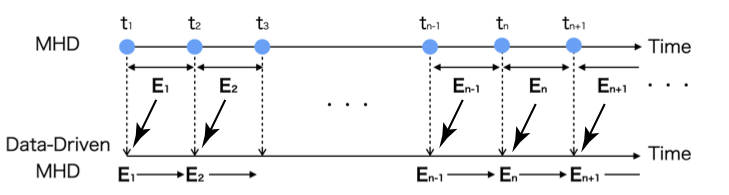}
  \caption{        
                This cartoon shows how advances in time are handeled by the data-driven MHD simulation. The blue circle corresponds to the ground-truth 
                data. We calculate the electric field $E_n$ at the bottom surface from the magnetic fields $B_n$ and $B_{n+1}$ at $t=n$ and $t=n+1$ in the   
                ground-truth data where $n=1,2,3, \cdots$.  In the data-driven simulation, $E_n$ drives an entire magnetic field in a time range from 
                $t_{n}$ to $t_{n+1}$. In this study, the time interval $t_{n+1}-t_n$ is set to 0.5625.
          }
  \label{f3}
  \end{figure}

    The coronal magnetic fields is driven as outlines in  Fig. \ref{f3}. We derive the electric field $\vec{E}^{(1)-(3)}$  at each time interval from the two output data at $t_n$ and 
    $t_{n+1}$ ($n$=1,2,3$\cdots$) obtained from the referenced MHD simulation where only the magnetic field at the bottom is used. The time interval $t_{n+1}-t_n$ is set  to 0.5625. 
    The electric fields are derived from the three Poisson equations (\ref{poisson_1}), (\ref{poisson_2}) and (\ref{poisson_3}). In this study,  we use the Gauss-Seidel method to solve them. 
    The performance and convergence of the Poisson equations are shown in Appendix (see Fig.\ref{fA_1}). In the data-driven simulation,  each electric field drives the upper coronal magnetic 
    field at  each interval. Note that, in this study,  we drive the coronal magnetic field when the $|\vec{B}|=\sqrt{B_x^2+B_y^2+B_z^2}$ measured at the bottom is more than 0.0625, {\it i.e.}, the 
    weak magnetic field region does not change with time.
        
      As seen in Table-1, we ran 8 different simulations which are mainly classified into the cases of TW and the cases of CV where TW and CV mean data-driven simulations in twisting 
    (phase energy buildup phase: $t$=0$\sim$11.25 in the MHD simulation)  and the converging phase  (energy released phase: t$\ge$22.5 in the MHD simulation), respectively. The 
    initial conditions of the data-driven MHD simulations in the cases of TW and CV employ the magnetic fields at $t$=0 and $t$=22.5 obtained from the ground-truth data. Note that the 
    data-driven simulations are not conducted continuously from $t$=0 to after the eruption. Because one of objectives of this study is that how the data-driven simulation can produce the 
    energy release phase. If we conduct the data-driven simulation from t=0 to the eruption, it becomes difficult to isolate whether the errors are due to the energy buildup phase (t=0-11.5) or 
    the relaxation  phase (t=11.5-22.5). We focus on evaluating the data-driven simulation in the energy buildup phase without the extra errors.  The Type-A  and -B differential methods denote -A 
    (A0, A1, and A2 ), and  -B, respectively. The Type-A is further classified into A0, A1, and A2 where the vertical velocity ($v_z$) placed at one grid ($k=1$) above the bottom surface has 
    been given through a specific update in the temporal evolution and the method is different in each case. In the cases of A1 and A2, the vertical velocity ($v_z$) is fixed to zero or is update 
    through a linear interpolation with values of  $k=0$ and $k=2$, respectively, while the A0 has no specific updated, {\it i.e.}, the vertical velocity is derived from an equation of motion directly. 
    In case B, we run one simulation where the vertical velocity at $k=1$ is updated using the liner interpolation.

    \section{Results}
    \subsection{Overview of Ground-truth Data}
\begin{figure}
  \epsscale{.9}
  \plotone{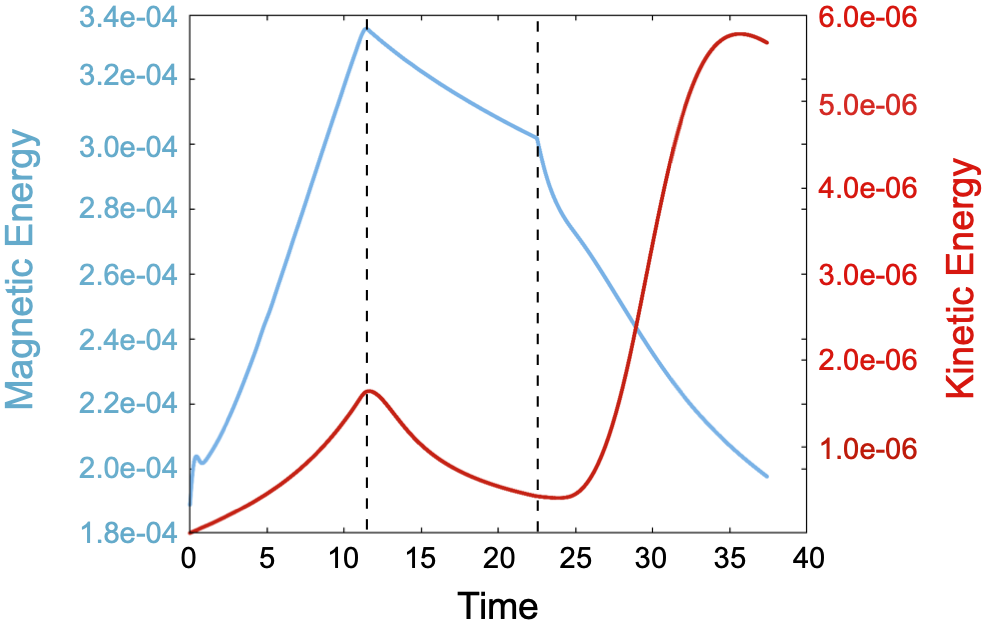}
  \caption{
               Temporal evolutions of the magnetic energy and the kinetic energy are plotted in blue and red, respectively. These are driven by the twisting 
               and diverging motions on the photosphere given in the MHD simulation. The vertical dashed lines are set at $t \sim$11.25 and $t \sim$22.5, 
               respectively. The twisting motion as shown in Fig.\ref{f1}c is imposed during a period, 0 $\le t \le$11.25 and after that the magnetic field is relaxed, 
               {\it i.e.}, no external velocity is imposed by $t \sim$22.5. After $t \sim$ 22.5, the converging motion is added on the bottom surface as shown in 
               Fig. \ref{f1}d.
               }
  \label{f4}
  \end{figure}

    The temporal evolution of the magnetic and kinetic energies are shown in Fig.\ref{f4}. Both energies are buildup by $t \sim$ 11.25 due to the twisting motion of the sunspot,  
    then gradually decrease by $t \sim$ 22.5 in the relaxation phase. Note that the twisting motion is halted at $t \sim$ 11.25 and the magnetic fields are relaxed by 
    $t \sim 22.5$. Within a few moments of turning on the converging motion  at $t \sim$ 22.5, the kinetic energy increases dramatically, whilst the magnetic energy decreases
     a lot of which is dissipated and converted into the Joule heating through the flux cancellation.
         
    The evolution of  the 3D magnetic field lines, while the twisting motion is imposed, is shown in Fig.\ref{f5}. As time passes, the magnetic field lines are sheared gradually and 
    eventually a sigmoidal structure is formed at $t$=10.12 which is often observed prior to flares ({\it e.g.}, \citealt{2012ApJ...750...15S}, \citealt{2012ApJ...747...65I}, \citealt{2018ApJ...869...99K}).
    
\begin{figure}
  \epsscale{1.}
  \plotone{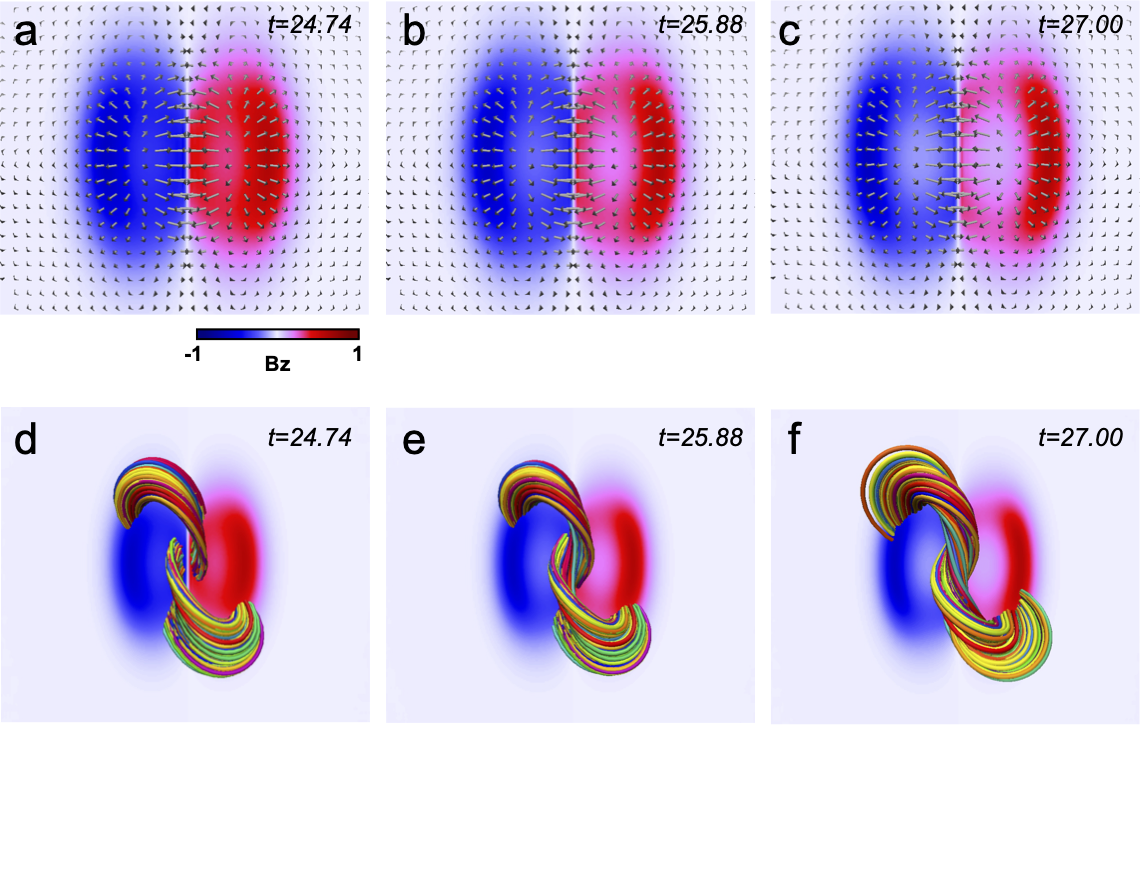}
  \caption{
               {\bf a-c} Temporal evolution of the converging motion given on the bipolar fields and $B_z$ component in color. 
               {\bf d-f} Temporal evolution of the magnetic field lines. In this period, the highly twisted lines (MFR) are being formed, which 
                            transitions from the pre-erupting stage to the erupting stage.
               }
  \label{f6}
  \end{figure}

\begin{figure}
  \epsscale{1.}
  \plotone{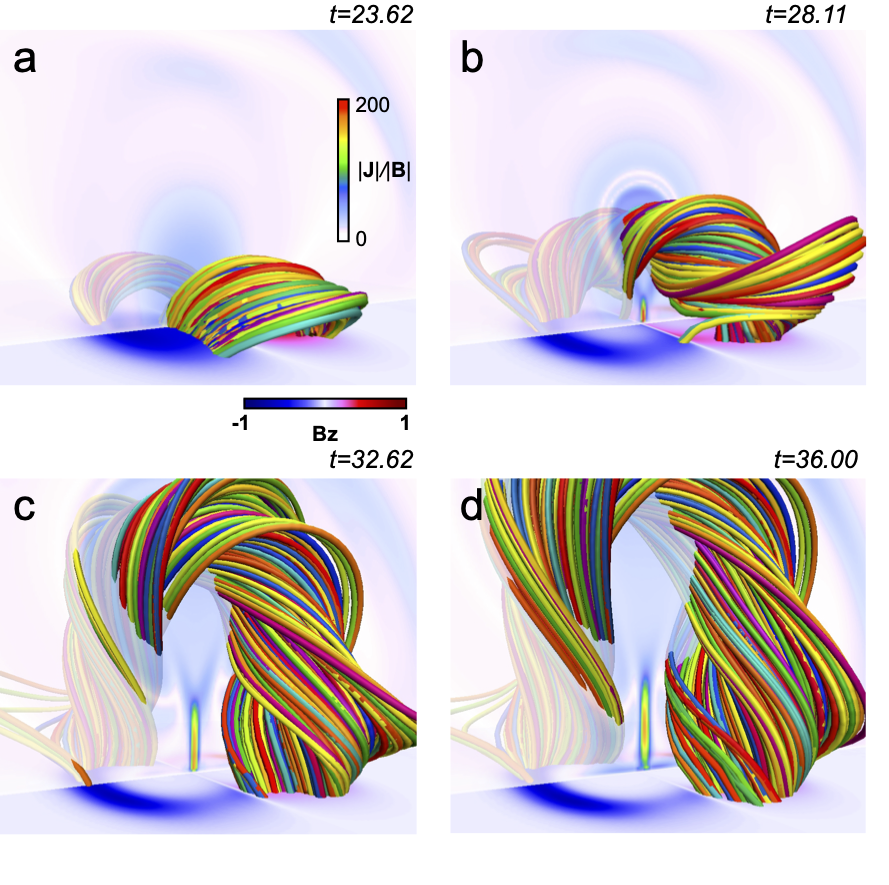}
  \caption{
               {\bf a-d}  Temporal evolution of the 3D magnetic field associated with the converging motion imposed on the bipolar field 
               given at the bottom surface. The lines correspond to the magnetic field lines and the value of $|\vec{J}|/|\vec{B}|$ is plotted 
               on the vertical cross section.  The bottom cross section is plotted in the $B_z$ distribution.
               }
  \label{f7}
  \end{figure}

    After the relaxation, we impose the diverging flow on the sunspots as shown in Figs.\ref{f6}a-c with the evolution of the sunspots. From the upper panels, a part of the magnetic 
    flux is transported toward the polarity inversion line (PIL) at which they are canceled.  The lower panels (Figs.\ref{f6}d-e) show the evolution of the magnetic field lines. The footpoints 
    of the magnetic field lines are carried by the flow on the both sunspots and field lines of opposite polarity encounter each other at the PIL. Magnetic reconnection then
    takes place, resulting in the formation of the long and highly twisted lines as seen in Figs.\ref{f6}e and f.  The continuous converging flow around the PIL enhances the further twisted MFR 
    and lifts it upward as seen in Fig.\ref{f7}b where the magnetic field lines just before the diverging flow are imposed are shown in Fig.\ref{f7}a. The current sheet is formed under the lifted 
    MFR and eventually it erupts as shown in Figs.\ref{f7}c and d. 
    
    We have attempted to reproduce these MHD processes by using the data-driven MHD simulation in which the electric field works as the driving source on the bottom surface, instead 
    of the velocity field. These electric fields are derived from a time series of the magnetic field on the bottom surface of the ground-truth data produced  by the MHD simulation.
   
    \subsection{Results of the Data-driven MHD Simulation}
   \subsubsection{Energy Buildup Phase}
\begin{figure}
  \epsscale{1.}
  \plotone{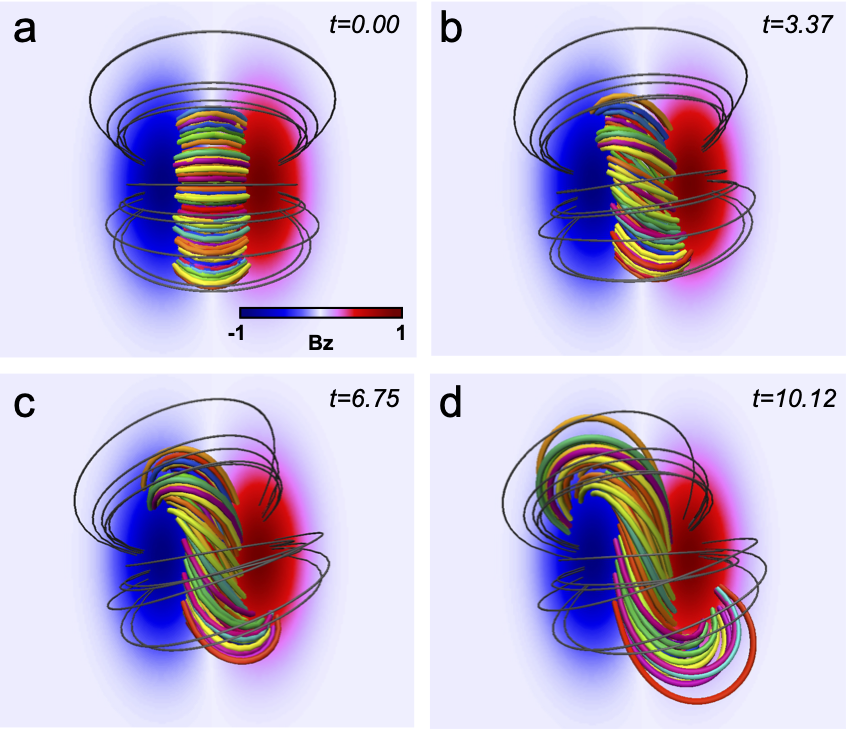}
  \caption{
               {\bf a-b}  Temporal evolution of the 3D magnetic field lines associated with the twisting motion imposed on the bipolar field given 
               at the bottom surface. The lines correspond to the magnetic field lines: colored lines are twisted and the black lines are overlying 
               field lines. As time passes, an S-shaped structure is being gradually formed.      
               }
  \label{f5}
  \end{figure}

\begin{figure}
  \epsscale{1.}
  \plotone{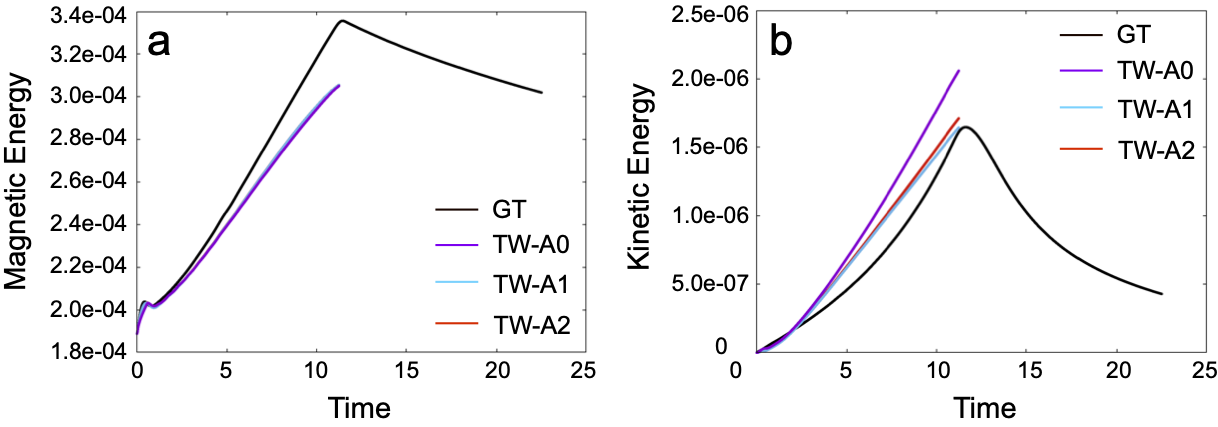}
  \caption{
                {\bf a} Temporal evolution of the magnetic energy in the energy build-up phase where the black, purple, light blue and red,  
                           represent the results in the ground-truth data and each case of the data-driven MHD simulation, TW-A0, TW-A1, 
                           and TW-A2, respectively. These lines obtained from the data-driven simulations are almost overlapping. 
                {\bf b} Temporal evolution of the kinetic energy. The format is the same as in {\bf a}.
               }
  \label{f8}
  \end{figure}

    First, we show the results of data-driven simulation in energy buildup phase in which the ground-truth data shows the formation of Sigmoid. The initial condition of 
   the data-driven simulation was given by the potential magnetic field shown in Fig.\ref{f1}b and the time-dependent electric fields are given at bottom and one-half the above (and also below).
   Figures \ref{f8}a and b show the temporal evolution of the magnetic and kinetic energies, respectively.  From Fig.\ref{f8}a, we can see that the temporal evolutions of magnetic energies 
   in the data-driven simulations are almost  indistinguishable to each other. It should also be noted that the quantities from the data-driven simulations are very similar to the initial ground-truth 
   data, but as the evolution continuous they deviate from this. Note that free magnetic energy is essential  to discuss the eruption rather than the net magnetic energy shown in this study. 
   However, the  potential field is exactly same in the ground-truth and data-driven simulations because the bottom $B_z$ is exactly same between them. Therefore, the profile of the free magnetic 
   energy is same to the one shown in Fig.\ref{f5}(a) while the magnitudes differ between the net magnetic energy and free magnetic energy.   From Fig.\ref{f8}b, although the kinetic energy 
   increases in the temporal evolution in both the ground-truth data and each data-driven simulation, the behaviors are slightly different among them. We note that the kinetic energy is much smaller 
   than the magnetic energy.
   
\begin{figure}
  \epsscale{1.}
  \plotone{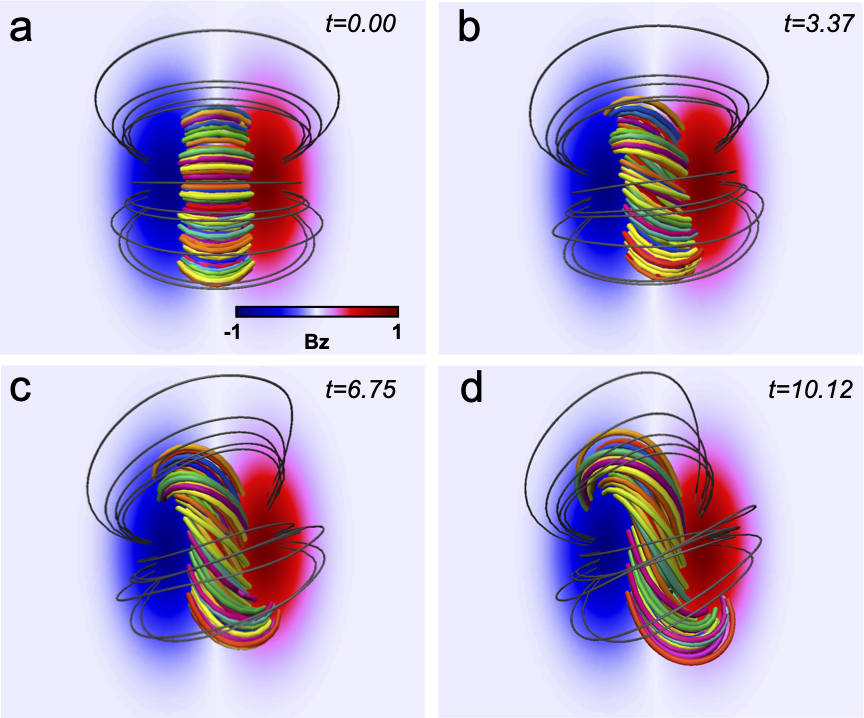}
  \caption{
               {\bf a-d}  Temporal evolution of the 3D magnetic field lines obtained from the data-driven MHD simulation (TW-A2).
                              The format of this figure is quite same as the one in Fig.\ref{f5}.        
               }
  \label{f9}
  \end{figure}

   The temporal evolution of the 3D magnetic field obtained from the data-driven simulation (TW-A02) is shown in Fig.\ref{f9} . The data-driven simulation reproduces a similar evolution to the 
   ground-truth data as shown in Fig.\ref{f5}. In particular, the sigmoidal structure is reproduced well in the final time-step. On the other hand, the sheared field lines accumulated in the 
   data-driven simulation look weaker than those in the ground-truth data. Nevertheless, this result shows that the data-driven simulation works well.   
     
\begin{figure}
  \epsscale{1.}
  \plotone{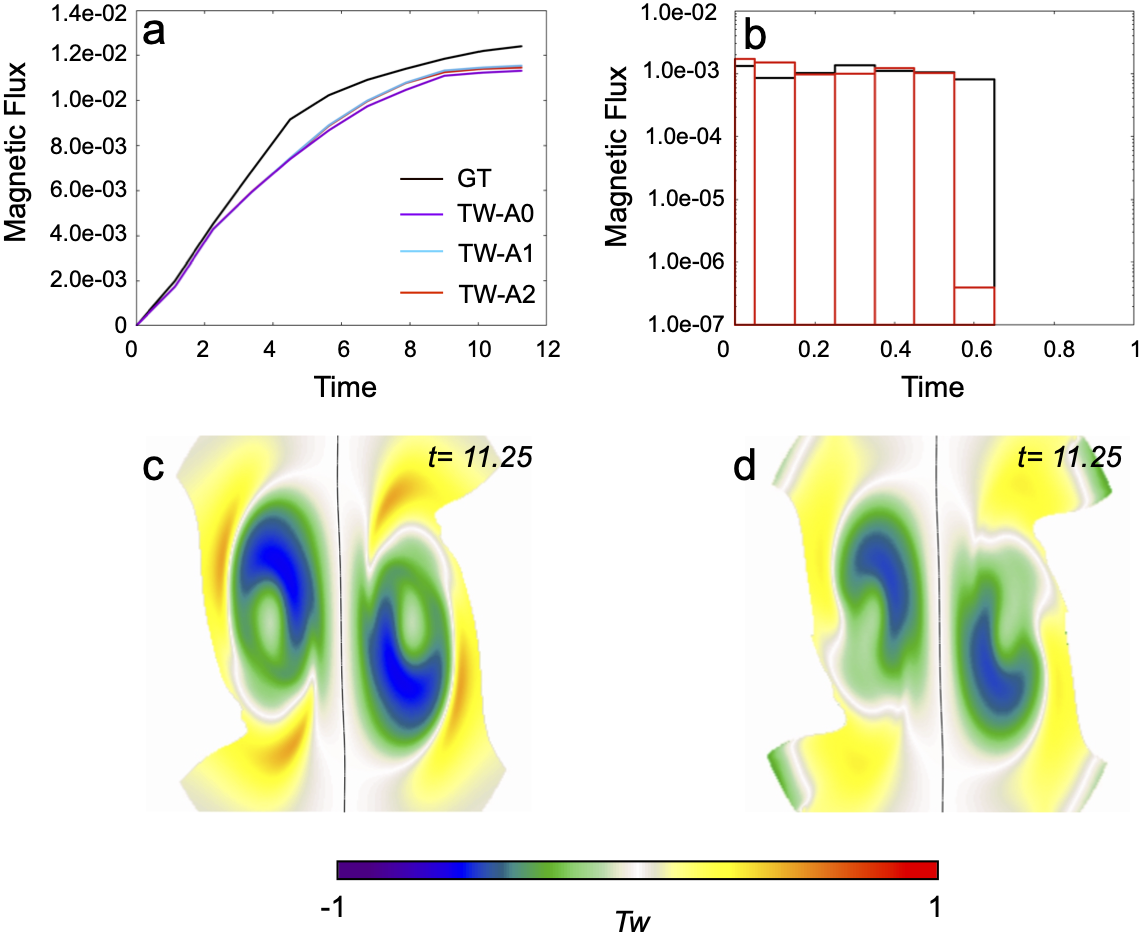}
  \caption{
               {\bf a} Temporal evolution of the magnetic flux dominated by the magnetic field lines which satisfy $T_w \le -0.1$ where the field 
                          lines are traced from the surface one grid above the bottom surface to measure the twist. The each colored line is same 
                          format as the ones in Fig.\ref{f8}.  
               {\bf b} Histogram for the magnetic flux vs. $T_w$ from the ground-truth data in black and the data-driven simulation in red (TW-A2), 
                          respectively. These results are obtained at $t$=11.25 which is the last time step in the both simulations.  
               {\bf c-d}  The snapshots of the twist distribution of each field line, which are mapped on the each bottom surface, at $t$=11.25 in 
                             the ground-truth data and in the data-driven MHD simulation (TW-A2). The black line corresponds to the PIL. Strong negative 
                              twist regions correspond to the footpoints of the MFR.
                   }
  \label{f10}
  \end{figure}

   We show more quantitative results by using magnetic twist defined as 
   \begin{equation}
   T_w = \int \frac{\vec{\nabla}\times \vec{B}\cdot\vec{B}}{|\vec{B}|^2} dl, 
   \end{equation}
   where $dl$ is a line element of each field lines (\citealt{2006JPhA...39.8321B}). Fig.\ref{f10}a shows the temporal evolution of the magnetic flux in the ground-truth data and each case 
   of the data-driven simulation. These magnetic flux are stored by the magnetic field lines which satisfy a condition $T_w \le -0.1$. Although the magnetic flux, which is composed of 
   the non-potential component, obtained from the data-driven simulation is a bit smaller than the magnetic flux obtained from the MHD simulation, the behavior is almost same among them. 
   Fig.\ref{f10}b shows the histogram of the magnetic flux which depends on the twist value for the ground-truth and data-driven simulation (TW-A02). Although the twist value obtained from 
   the data-driven simulation is slightly weaker than that calculated from the ground-truth data, the two distributions are very similar. The distributions of the magnetic twist of each field line 
   are mapped on the each surface.  The results calculated from the ground-truth data and the data-driven MHD (TW-A02) simulation are also very similar as seen in Figs.\ref{f10}c and d. 
   Thus, these results support that the data-driven simulation reproduces the twisting process of the magnetic field lines well. 
  
      \subsubsection{Energy Release Phase}
\begin{figure}
  \epsscale{1.}
  \plotone{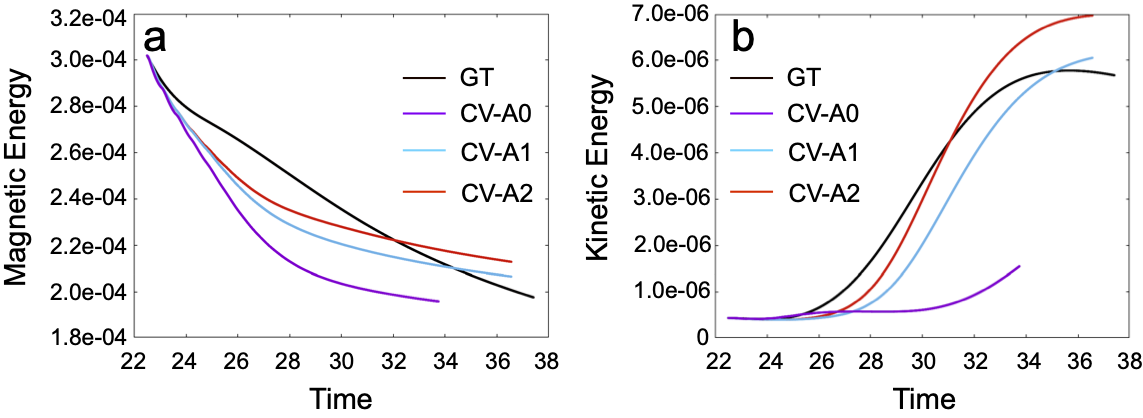}
  \caption{
                {\bf a} Temporal evolution of the magnetic energies in the period from the pre-erupting to the erupting phase. The black, purple, 
                          light blue, and red  lines are the results obtained from the ground-truth data and the data-driven MHD simulations 
                          (CV-A0, CV-A1, and CV-A2), respectively.
                {\bf b} The temporal evolution of the kinetic energy whose format is the same as in \bf{a}.
               }
  \label{f11}
  \end{figure}

    Next we show the results obtained from the data-driven MHD simulation in the energy release phase. This corresponds to the erupting phase of the MFR formed by the converging 
    motion. The initial condition of the data-driven simulation was given by the magnetic field at $t = 22.5$ of the ground-truth data. Figs.\ref{f11}a and b show the temporal evolutions 
    of the magnetic and kinetic energies obtained from the ground truth and data-driven simulations. We can see that the magnetic and kinetic energies obtained from the data driven 
    simulations capture the tendency to decrease and increase as is seen in the respective ground truth data. However, in the case CV-A0 the kinetic energy grows very slowly compared 
    to other cases, which will be discussed later. 
   
     As seen in Fig.\ref{fA_3}, the bottom $B_z$ obtained from the data-driven simulation reproduces the ground-truth data well. The correlation coefficient is over 0.99 
    at each time. This means that the potential fields extrapolated from the bottom $B_z$ of the ground-truth data and obtained from the data-driven simulation are almost  identical. 
    Although the magnitudes of the net magnetic energy and free magnetic energy are different from each other, the behavior is the same as that shown in Fig.\ref{f11}(a). 
       
\begin{figure}
  \epsscale{1.}
  \plotone{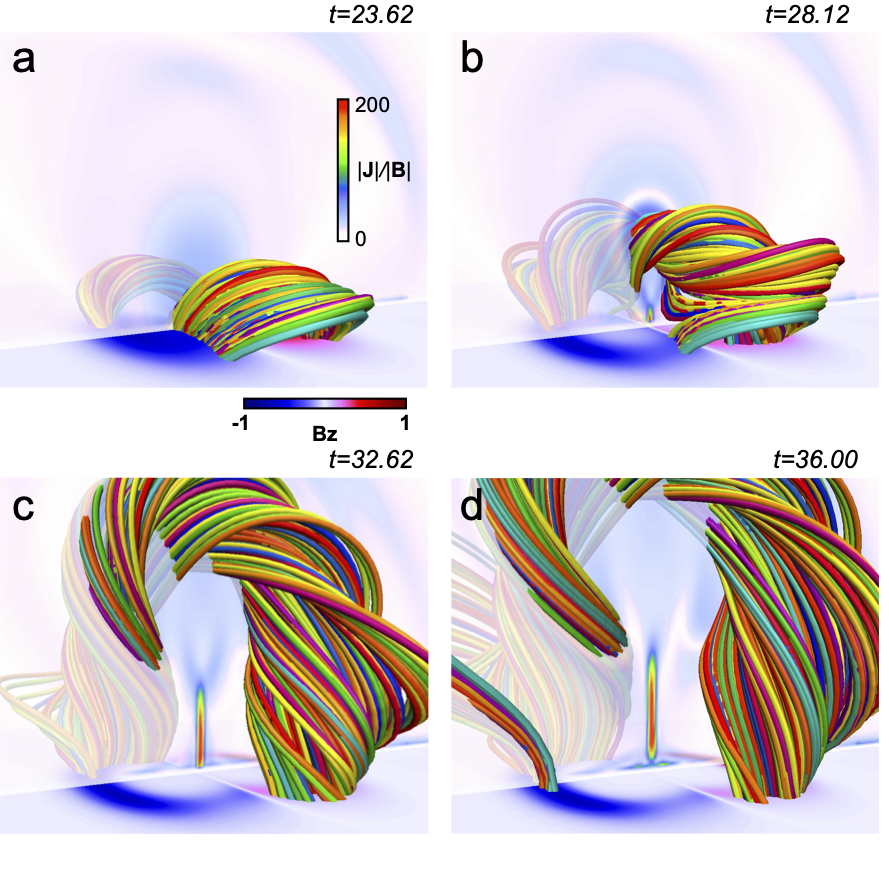}
  \caption{
               {\bf a-d}  Temporal evolution of the 3D magnetic field obtained from the data-driven MHD simulation (CV-A2).  These formats
                              are the same as  those in Fig.\ref{f7}.     
                }
  \label{f12}
  \end{figure}

    The temporal evolution of the 3D magnetic field lines in the case of CV-A2 is  shown in Fig.\ref{f12}. The MFR is formed through the converging motion which is driven by the electric field, 
    and causes an eruption, under which the current sheet is enhanced. We confirmed that the evolving 3D magnetic field shown in the data-driven simulation is consistent with the ground-truth 
    data, thus the data-driven simulation looks to work well in the energy release process of the magnetic field that corresponds to the erupting phase of the MFR.   
    
\begin{figure}
  \epsscale{1.}
  \plotone{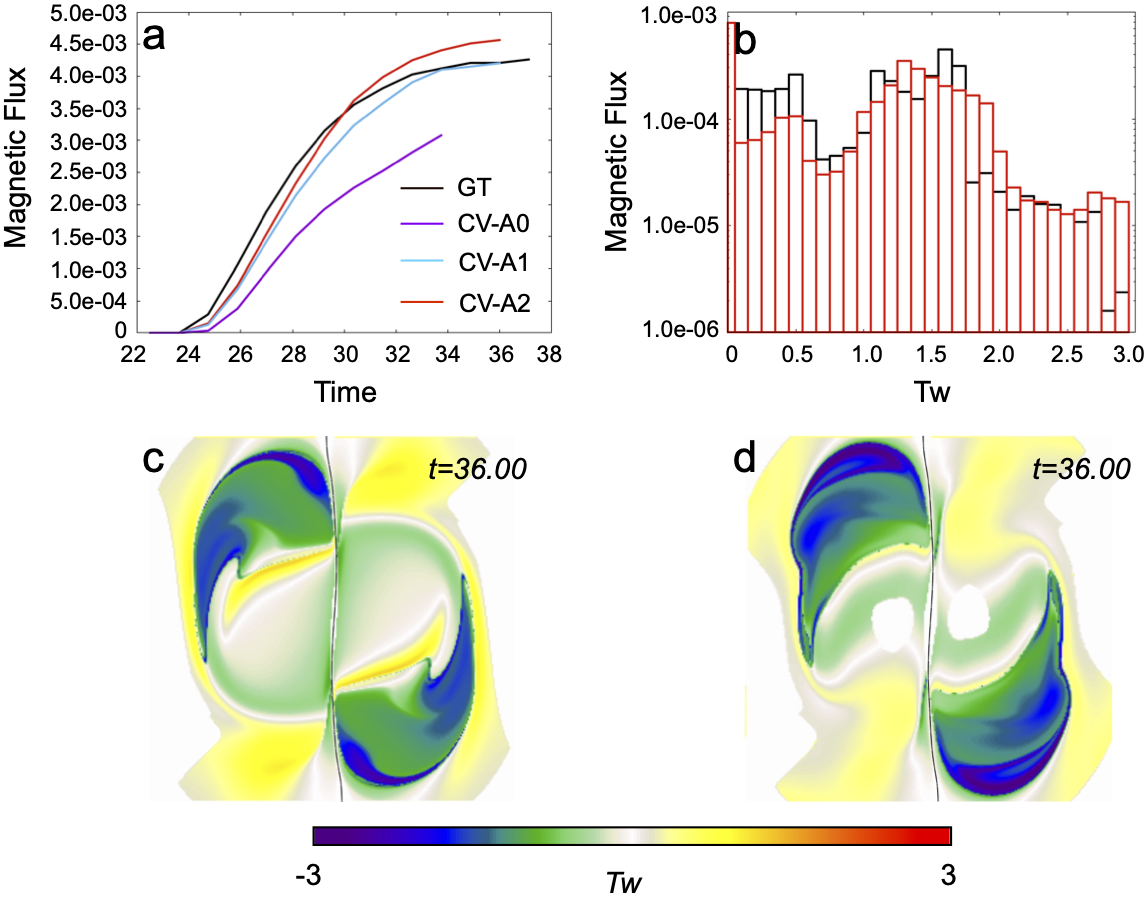}
  \caption{
               {\bf a} Temporal evolution of the magnetic flux dominated by the highly twisted field lines which satisfy $T_w \le -1.0$. These highly 
                          twisted field lines are newly created during the eruption.  Each colored line represents the same format as in  Fig.\ref{f11}(a).              
               {\bf b} Histogram for the magnetic flux vs. $T_w$ where the black and red lines represent the results obtained from the ground-truth 
                         data and the data-driven simulation (CV-A2) at $t$=36.00, respectively. These are results at the last time of the each 
                         simulation. 
               {\bf c-d} Snapshots of the twist distribution of each field line, which are mapped on the each surface, obtained from the ground-truth 
                             data and the data-driven simulation (CV-A2) at $t$=36.00, respectively.  The format is the same as in Figs.\ref{f10}c and d.
                }
  \label{f13}
  \end{figure}

    The results of the quantitative analysis (especially focusing on the case of CV-A2) are shown in Fig.\ref{f13}. Figure \ref{f13}a shows the temporal evolution of the magnetic flux, which is 
    dominated by the highly twisted field lines, $|T_w| \ge 1.0$, in each case where these highly twisted field lines are newly created during the evolution. The evolution obtained from 
    the data-driven simulation in each case almost captures the ground-truth data (black) while the growth in the case of CV-A0 is very slow as inferred from the result 
    shown in Fig.\ref{f11}b. Figure \ref{f13}b represents a histogram for the magnetic flux vs. magnetic twist for the ground-truth data and data-driven simulation (CV-A2). Figs.\ref{f13}c 
    and d show the distribution of the $T_w$ mapped on the surface, obtained from the ground-truth data and the data-driven simulation (CV-A2), respectively. Although these do not 
    match exactly, both are very similar. Thus we confirmed that the data-driven simulation works well in terms of the quantitative analysis.
    
\begin{figure}
  \epsscale{.9}
  \plotone{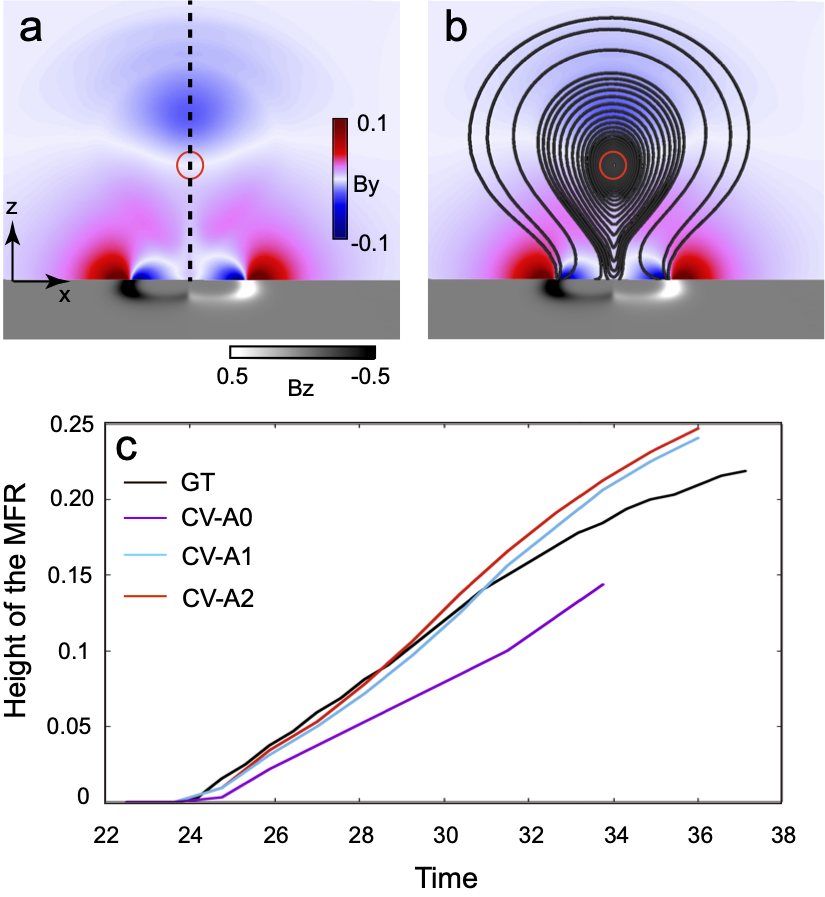}
  \caption{
               {\bf a} The $B_x$ distribution obtained in the ground-truth data is plotted on the vertical cross section at $y=0.5$ at $t$=34.87. 
                         The vertical dashed line is plotted from the position ($x,y$)=(0.5,0.5) and  the red circle is located at $B_x$=0.
               {\bf b} The field lines, which are traced in $x-z$ plane, are plotted over (a). The red circle, which is the same one in {\bf a}, points 
                          out the center of the MFR.
               {\bf c} The temporal evolutions of height profile of each MFR calculated from the ground-truth and the each data-driven simulation.
                          The color is the same format as in Fig.\ref{f11} or Fig.\ref{f13}a.
                }
  \label{f14}
  \end{figure}

    We trace the magnetic axis of the MFR in the evolution for the ground-truth data and the data-driven MHD simulations, respectively. In this study, we detect the axis at which the 
    sign of $B_x$ inverts along the center of the numerical box which corresponds to the dashed vertical line as shown in Fig.\ref{f14}a. Because the symmetry of the MFR is well during the 
    evolution. Figure \ref{f14}b plots the field lines on the vertical cross section (x-z plane) and the red circle points out the same position shown in Fig.\ref{f14}a which corresponds to a center 
    of the MFR. The temporal evolutions of the MFR axis of the ground-truth data and those obtained from each case of the data-driven simulation are shown in Fig.\ref{f14}c. Although case 
    CV-A0 deviates from the result from the ground-truth data, other cases obtained from the data-driven simulations capture its behavior well. Therefore, the data-driven simulation developed 
    in this study, which is based on \cite{2018ApJ...855...11H}, can be used as a powerful tool to understand the evolution of the coronal magnetic field  and the physics of  solar eruptions.
  
    \section{Discussion}
  \subsection{Why is a special update required for the vertical velocity located at $z=\Delta z$?}
    In above section, we found that the case of CV-A0 does not capture the ground-truth data well, for instance, in the temporal evolution of the kinetic energy and the axis of the 
    MFR. One other notable difference when compared to the other cases from the data-driven simulation is that the eruption occurs later in the CV-A0 case. The difference between 
    these cases is the handling of the vertical velocity ($v_z$) at $k=1$, {\it i.e.}, $z=\Delta z$. Therefore, we should examine the behavior of the vertical velocity at $z=\Delta z$ or 
    around the position, in both energy build up and energy release phases, respectively. 
    
\begin{figure}
  \epsscale{1.}
  \plotone{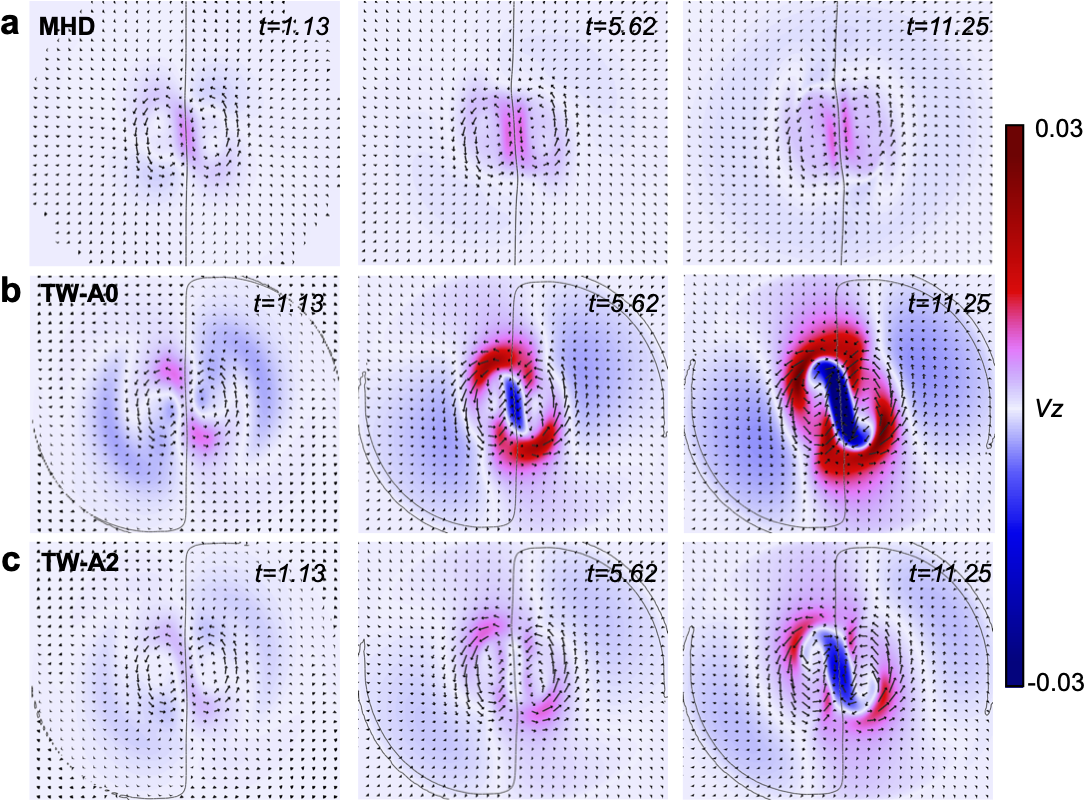}
  \caption{Temporal evolutions of the velocity at a surface at $z=3\Delta$ obtained from the MHD simulation and the results of the 
                data-driven MHD simulations in the cases of TW-A0, and TW-A2. The color shows the $v_z$ distribution while the arrow 
                indicates horizontal velocity $\vec{v}_h=(v_x, v_y)$ and the black line corresponds to the PIL.}
  \label{f15}
  \end{figure}

    Figure \ref{f15} shows the temporal evolutions of the velocity field obtained from the ground-truth data and the data-driven MHD simulation (TW-A0 and TW-A2), respectively, 
    at the surface at $z=3\Delta z$. Although the horizontal velocity plotted by the arrows forms a twisting motion in each case, the vertical velocity ($v_z$) distributions are quite 
    different. The most striking difference is that the negative vertical velocity appears in late phase of the data-driven MHD simulations. So, we hereafter discuss the spacial and temporal 
    evolutions of the vertical velocity.  
    
\begin{figure}
  \epsscale{1.}
  \plotone{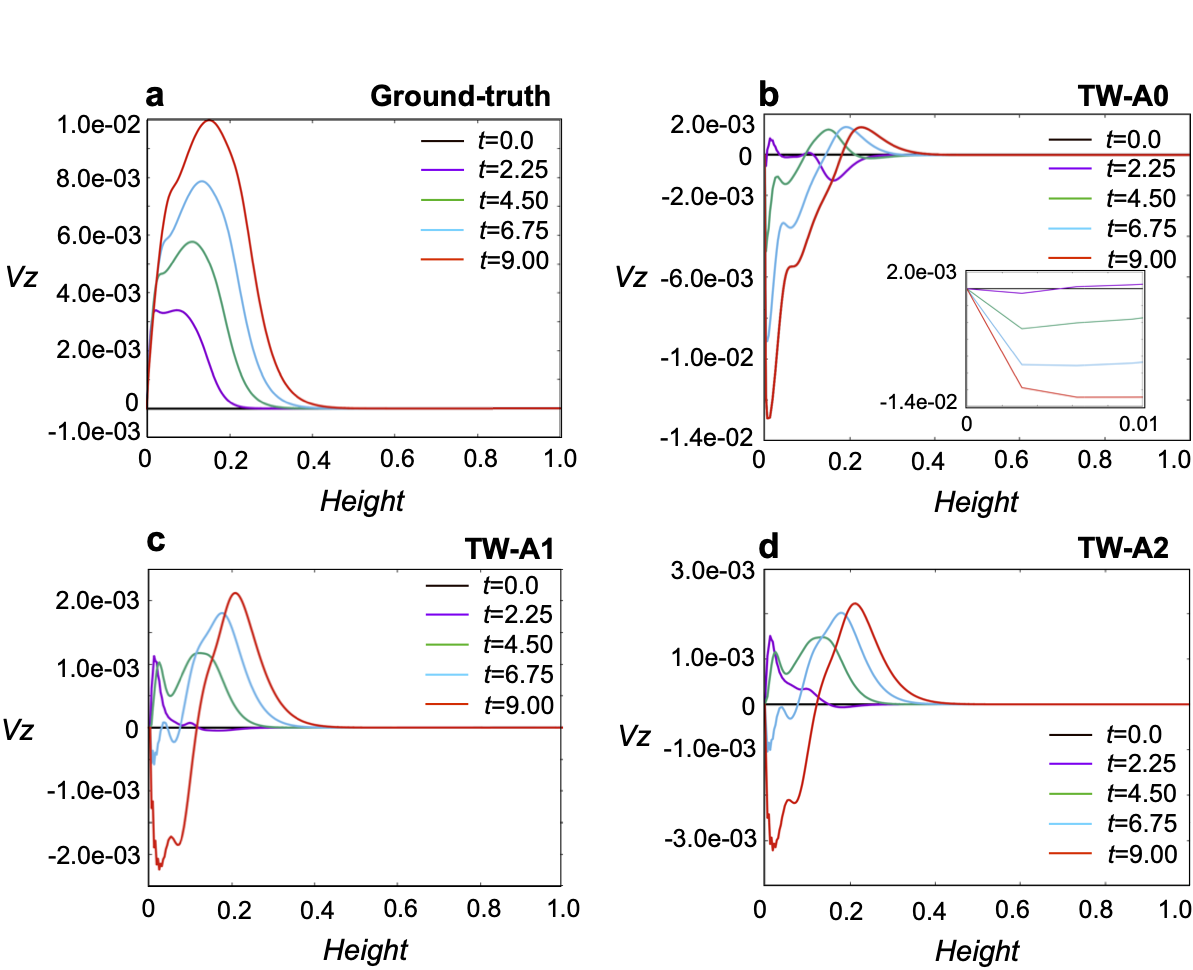}
  \caption{The temporal evolution of the $v_z$ at a center of the  numerical box which corresponds to $\vec{r}_c=$(0.5, 0.5, z). The vertical and horizontal axes 
                 correspond to $v_z$ and height (z), respectively. These velocities are associated with the twisting motion given on the photosphere. {\bf a-d} correspond 
                 to the results obtained from the ground-truth data and the data-driven simulations (TW-A0, TW-A1, and TW-A2). The each colored line represents 
                 the velocity profile in z-direction at each time where the black, purple, green, blue, and red lines correspond to each time, $t=$0.0, 2.25, 4.50, 6.75, and 9.00, 
                respectively. The small inset corresponds to the enlarged view in the height range from 0 to 0.01. }
  \label{f16}
  \end{figure}

    Figure \ref{f16}  shows the temporal evolutions of the vertical velocity along the center of the numerical box, {\it i.e.}, the vertical dashed line as shown in Fig.\ref{f14}a, which are obtained 
    from ground-truth data and  the data-driven simulations, respectively, in the energy buildup phase (TW-A0 to TW-A2). For the ground-truth data, the velocity temporally increases  according 
    to the twisting motion given on the photosphere. Although the vertical velocity in TW-A1 and TW-A2  increases as times goes on, it enhances the negative value with time in the region close 
    to the solar surface, which is different to the ground-truth data. The typical result is the case of TW-A0 in which the velocity is negatively increased close to the solar surface with time. The 
    small inset is an enlarged view in the height range 0 to 0.01. We found that the velocity located at one grid above the bottom surface, {\it i.e.}, $z=\Delta z$, is fastest way to enhance the 
    negative value.  The time variation of the vertical velocity in the case of TW-A0 is determined by the equation of motion while in TW-A1 it is set of zero and TW-A2 it is updated by a linear 
    interpolation in the vertical direction, respectively. Either the convective term $\vec{v}\cdot\vec{\nabla}\vec{v}$ or Lorentz force $\vec{J}\times\vec{B}$, or both would have a negative 
    influence on the vertical velocity. Despite each velocity profile begins different in each case of energy buildup phase, magnetic energy is found to follow almost the same behavior in each 
    case (Fig.\ref{f8}) . Since the magnetic energy is dominant over the kinetic energy, the evolution of the magnetic field is almost unaffected by the velocity field, conversely a small fluctuation 
    of the magnetic field would greatly influence the evolution of the velocity. Therefore, the Lorentz force would be a major factor inhibits the vertical velocity from leading the correct solution. 
    Note that, in the real situation, since no one can say that the magnetic energy dominates and the velocity field does not affect the magnetic evolution much on the photosphere, 
    the situation is  expected to become more complex.
    
\begin{figure}
  \epsscale{1.}
  \plotone{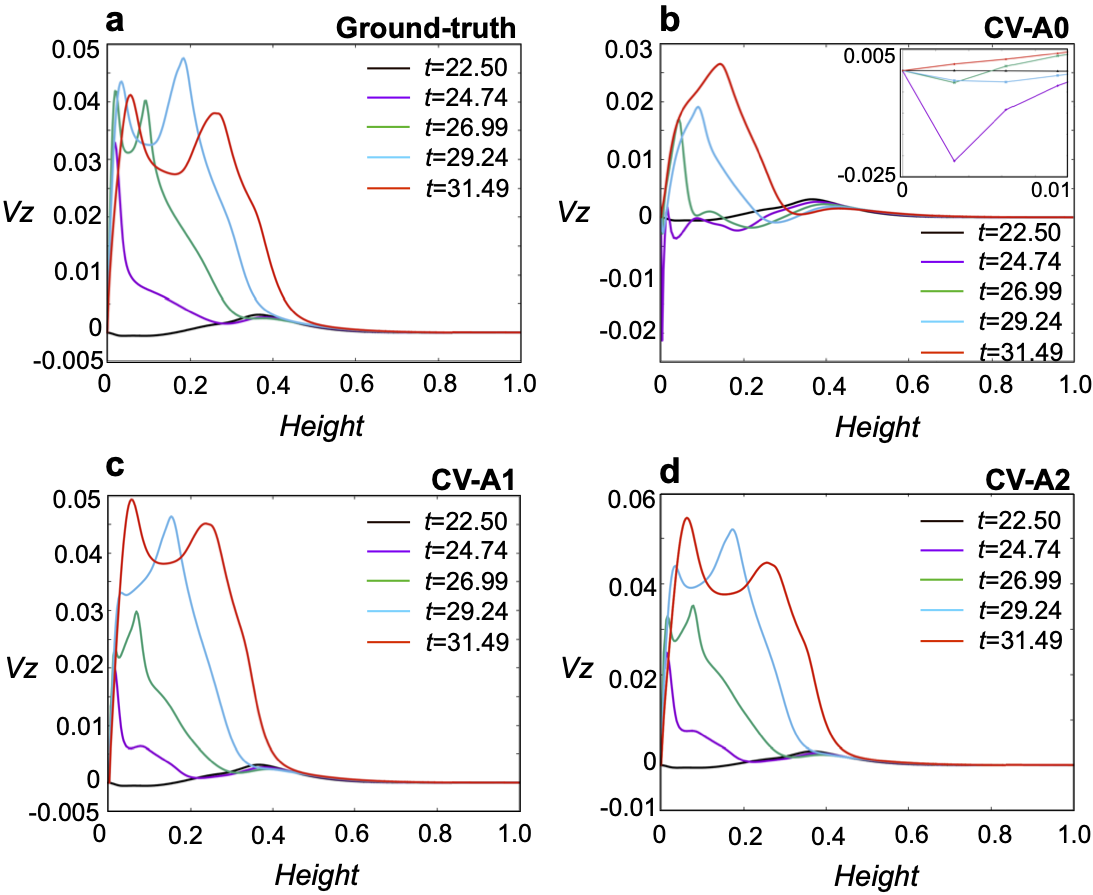}
  \caption{ The temporal evolution of $v_z$ at a center of the numerical box, associated with the converging motion given on the 
                 photosphere. The format is the same as Fig.\ref{f16} where {\bf a-d} correspond to the results obtained from the ground-truth 
                 data and the data-driven MHD simulations (CV-A0, CV-A1, and CV-A2). The black, purple, green, blue, and red lines 
                 represent the height profiles of the vertical velocities at $t$=22.50, 24.74, 26.99, 29.24, and 31.49, respectively. The small 
                 inset corresponds to the enlarged view in the height range from 0 to 0.01.
                }
  \label{f17}
  \end{figure}

    Figure \ref{f17} plots the temporal evolution of the vertical velocity in the energy release phase, {\it i.e.},  the erupting phase of the MFR in the same format as in Fig.\ref{f16}. The data-driven 
    simulations, CA-V1 and CA-V2 capture the ground-truth data well while the results obtained from CV-A0  show the different behavior.  From the small inset  follows the same format as in 
    Fig.\ref{f16}b, the value at $z=\Delta z$ suddenly becomes the negative value. This behavior is the same as seen in the case of TW-A0  in the energy buildup phase.  However, this case 
    differs from TW-A0 in that the velocity recovers from its negative value during the late phase.
        
\begin{figure}
  \epsscale{1.}
  \plotone{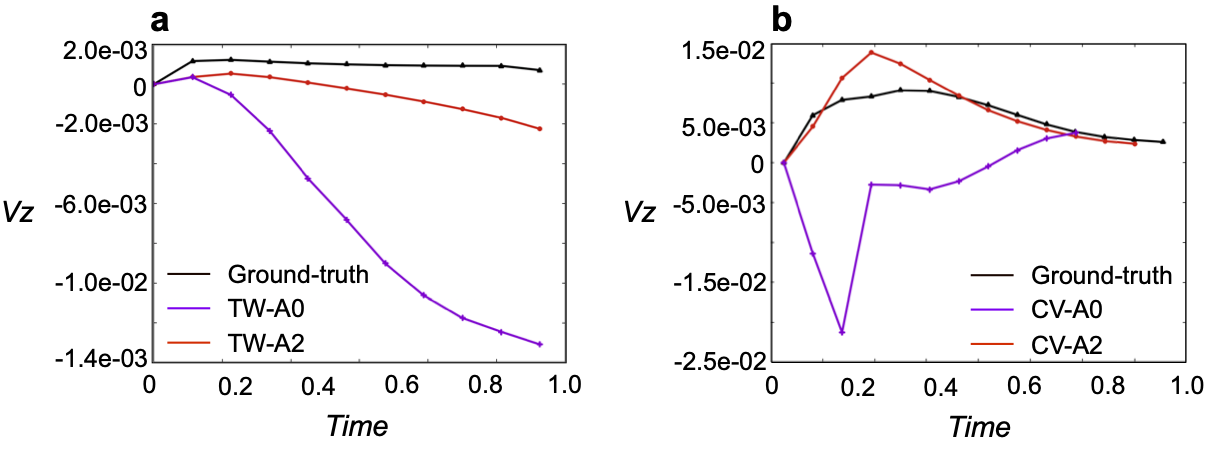}
  \caption{
               {\bf a} Temporal evolution of $v_z$  at $k$=1 is plotted for the each case in the energy buildup stage. Each color, black, 
                         purple and red, represents the results obtained from the ground-truth data and the data-driven simulations (TW-A0, and TW A2), 
                         respectively.
              {\bf b} Temporal evolution $v_z$ at k=1 is plotted for each case in the energy release stage. Each color, black, purple 
                        and red, represents the results obtained from the ground-truth data and the data-driven simulations (CV-A0 and CV-A2), 
                        respectively. }
  \label{f18}
  \end{figure}

    Figures \ref{f18}a and b compare the temporal evolutions of the vertical velocity located at $z=\Delta z$ for the ground-truth data and the results obtained from between the data-driven 
    simulation in energy buildup phase (TW-A0 and TW-A2) and energy release phase (CV-A0 and CV-A2). Note that  the vertical velocity at $z=\Delta z$ for the cases of  TW-A1 and CV-A1 
    is set to zero, so we exclude these plots. The TW-A2 and CV-A2 cases somewhat capture the ground-truth data, while cases TW-A0 and CV-A0 show large deviations from the ground-truth 
    by exhibiting a steep drop in the negative value from the initial values. CV-A0 however is found to increase the velocity back towards that seen in the ground-truth data. It is likely that when 
    the magnetic field is converted into the dynamics phase from the static phase, the strong positive enhancement of the velocity is associated with the eruption, which returns to the ground-truth 
    data even at $z=\Delta z$.    
    
    As seen in these results, we need a careful treatment of the vertical velocity at $z=\Delta z$ in the static phase, which is found to largely deviate from the ground-truth data without the 
    treatments. Since the kinetic energy in the energy buildup phase is much smaller than the magnetic energy, the evolution of the magnetic field is unaffected by the velocity field even if the 
    velocity strongly deviates from the ground-truth data as seen in Fig. \ref{f8}.  On the other hand, it affects the initiation of the eruption, which causes delays because the down flow, which is 
    an unlikely result in the ground-truth data, inhibits local reconnection at the photosphere and therefore the formation of the MFR. Thus, some treatments (simple linear interpolation was 
    applicable in this study) would be required.

    Why does the vertical velocity ($v_z$) located at $z=\Delta z$ show different behaviors in each case? From the above results, when the difference to the ground-truth data 
    is striking, the magnetic field 
    evolves in the quasi statical energy buildup phase toward the pre-erupting stage. Since the magnitude of the kinetic energy is much lower  than the magnetic energy in the 
    energy buildup phase,  $\vec{J}\times\vec{B}$ is the major factor to lowering velocity. Each component of the Lorentz force (\vec{F}) is described as 
    \begin{displaymath}
    F_x = J_yB_z - J_zB_y,
    \end{displaymath}     
    \begin{displaymath}
    F_y = J_zB_x - J_xB_z,
    \end{displaymath}
    \begin{displaymath}
    F_z = J_xB_y - J_yB_x,
    \end{displaymath}
    where $B_z$ at $z=\Delta z$ is derived from the rotation of the electric fields located at $z=\Delta z$ while a derivative of $E_y$ and $E_x$ in z component is included to 
    derive the $B_x$ and $B_y$, respectively,  according to equation (\ref{type-a}).  $B_z$ is determined by $E_{k=1}$ which is derived from MHD electric field 
    ($-\vec{v}\times\vec{B}$) while $B_x$ and $B_y$ are derived from $E_{k=2}$, $E_{k=1}$ and $E^{(3)}$  too (see Equation (\ref{type-a})). $E_{k=2}$ and $E_{k=1}$ are 
    different from $E^{(3)}$ because the $E^{(3)}$ is derived from a non-MHD process. Therefore, the z-derivative is generally not allowed at $z=\Delta z$ and the 
    physical consistency regarding of the obtained $B_x$ and $B_y$ is not guaranteed.  $F_z$ includes more  $B_x$ and $B_y$ than the other components $F_x$ and $F_y$, so 
    the value of $v_z$ that is derived from $F_z$ will have a lesser accuracy. 
        
  \subsection{What is difference between Type-A and Type-B?}
\begin{figure}
  \epsscale{1.}
  \plotone{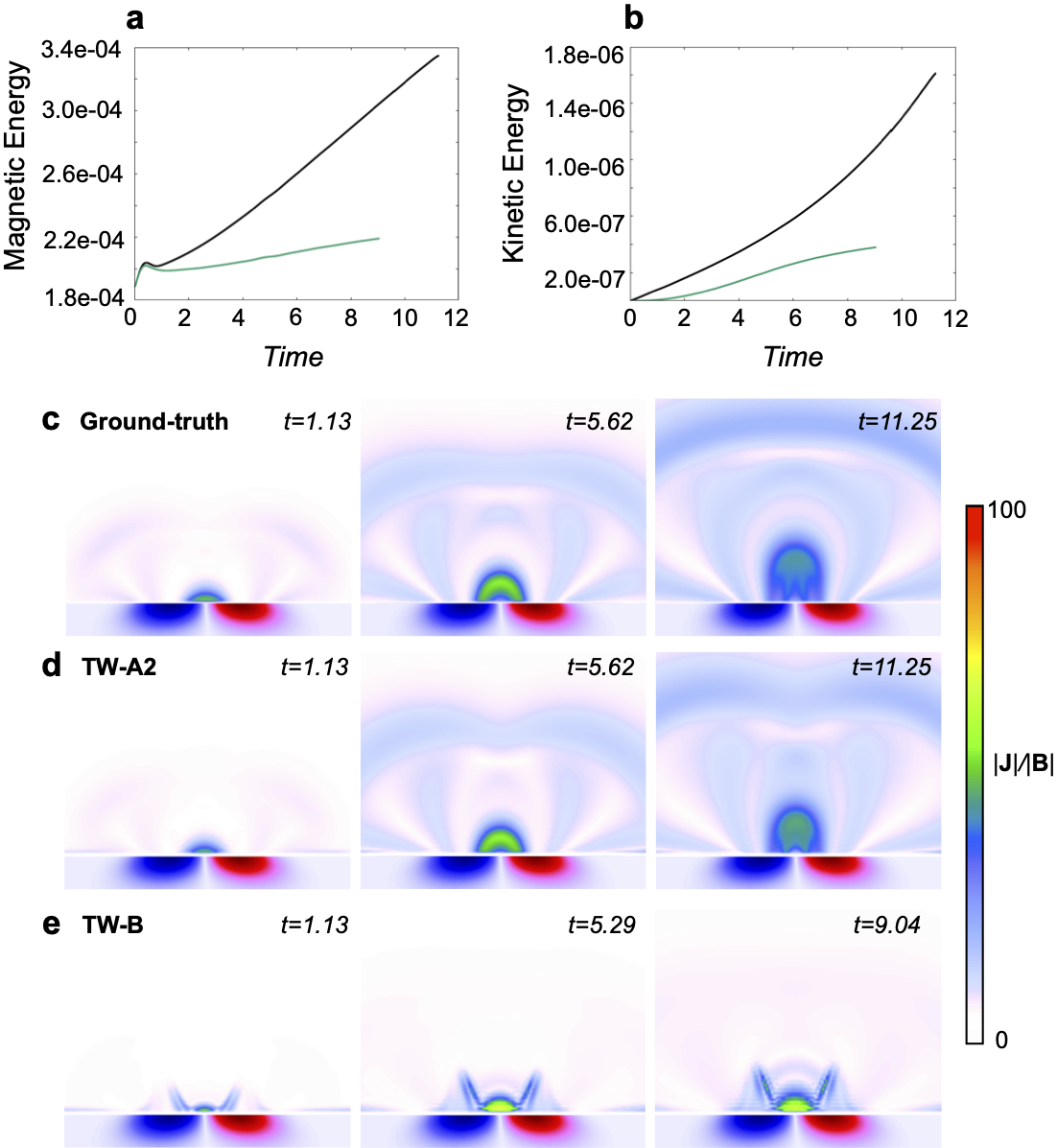}
  \caption{ {\bf a}      The temporal evolution of the magnetic energies obtained from the ground-truth data in black and the data-driven 
                                simulation (TW-B) in green.
                 {\bf b}     The temporal evolution of the kinetic energies in each case which are in the same format as {\bf a}.
                 {\bf c-e}  The temporal evolution of $|\vec{J}|/|\vec{B}|$ obtained from the ground-truth data and the data-driven simulations  
                               (TW-A2, and TW-B), respectively.}
  \label{f19}
  \end{figure}

    We now discuss results obtained from the data-driven simulation for the case of TW-B that implements a differential method type-B described in equation (\ref{type-b}). We first  
   compare with the results obtained from the ground-truth data. Figures.\ref{f19}a and b plot the temporal evolutions of the magnetic and kinetic energies in the energy buildup 
   phase, {\it i.e.}, the magnetic field is twisted due to the photospheric motion. Both evolutions obtained from the data-driven simulation shows large deviations from the ground-truth data. 
   Figures \ref{f19}c, d, and e show the distribution of $|\vec{J}|/|\vec{B}|$ plotted on the vertical cross section for the ground-truth data and the data-driven MHD simulations 
   (TW-A2 and TW-B), respectively,  from which the TW-A2 case reproduces the ground-truth data well, while the case of TW-B shows a totally different solution.
   
   The difference between the TW-A2 and TW-B is the application of differing numerical differential methods, especially, at $z=\Delta z$ at $k=1$. The derivative forms in the z direction 
   for each are as follows, with type-A written as  
   \begin{equation}
    \frac{\partial E}{\partial z}|_{k=1} =  \frac{1}{2}\left[\frac{E_{k=2} - E_{k=1}}{\Delta z}
                                                          + \left(\frac{E_{k=\frac{1}{2}} - E_0}{\frac{\Delta z}{2}}\right)\right],  
   \label{NEO-type-A}                                                       
   \end{equation} 
   and type-B is written as   
   \begin{equation}
   \frac{\partial E}{\partial z}|_{k=1} =  \frac{\frac{(E_{k=2} + E_{k=1})}{2} - E_{k=\frac{1}{2}}}{\Delta z}.   
   \label{NEO-type-B}
   \end{equation}   
 Type-A is an average value of the central differential method of electric fields obtained from the MHD process (the first term in the right hand side of Eq.(\ref{NEO-type-A})) and non-MHD 
  process (the second term). Although each derivative has physical meaning, there is a discrepancy between the MHD term and non-MHD term. The use of the average process in this 
  case smoothly connects them, making a more robust calculation. Type-B is a central differential method of the electric fields obtained from the MHD and non-MHD processes.  In principle, this 
  derivative is not allowed as discussed in above section because the origin of those electric fields is different, {\it i.e.}, there are not connected smoothly. Therefore, this discrepancy might 
  negatively impact the simulation result.
     
  On the other hand, \cite{2018ApJ...855...11H} and \cite{2019ApJ...871L..28H}  implement type-B and the simulation works well.  One reason for the difference in results from this 
  study is the difference in the numerical schemes used. \cite{2018ApJ...855...11H} and \cite{2019ApJ...871L..28H}  are designed with the finite volume method, and numerical 
  viscosity and resistivity might efficiently work on the negative impact and make the simulation stable. However, such viscosity and resistivity do not work efficiently in the central 
  differential method as used in this study despite these being included explicitly in the induction equation and the equation of motion. Therefore, numerical instability is not  suppressed 
  in this study and the solution is not reproduced correctly as seen in Fig.\ref{f19}e. 
  
  \subsection{Dependency on Time Cadence}
\begin{figure}
  \epsscale{1.}
  \plotone{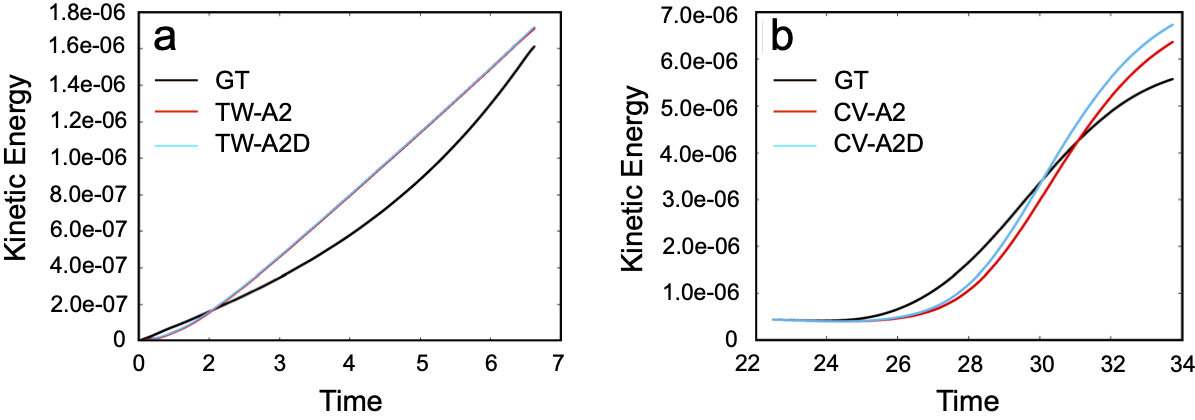}
  \caption{ {\bf a}      The temporal evolutions of the kinetic energy in the twisting phase for the ground-truth data in black, the 
                                data-driven simulations (TW-A2 and TW-A2D) shown in red and blue, respectively. TW-A2D has 2.5 times the 
                                temporal resolution of TW-A2. 
                 {\bf b}     The temporal evolutions of the kinetic energy in erupting phase for each case. The format is the same as those 
                                shown in {\bf a}.}
  \label{f20}
  \end{figure}

   We test the effect on the accuracy of our results to the time cadence used in the data-driven MHD simulations used in this study. Figure \ref{f20}a shows the comparison of the temporal 
  evolution of the kinetic energies in the energy buildup phase that is obtained form the ground-truth data, the data-driven simulation (TW-A2), and a new data-driven simulation with 2.5 
  times the temporal resolution ($t_{n+1}-t_n$=0.225) of TW-A2 ( TW-A2D). These are plotted in black, red, and blue, respectively. Since the blue line almost overlaps with 
  the red line, no significant differences due to the temporal evolution are apparent in the energy buildup phase. Thus we can say that the solution is convergent with respect to temporal 
  resolution.  

    Figure \ref{f20}b shows the each kinetic energy during the energy release phase, plotted in the same format as Fig. {\ref{f20}}a. The red line, which corresponds 
  to the data-driven simulation (CV-A2), diverges slightly from the blue line (CV-A2D) that has 2.5 times the temporal resolution of CV-A2. This is difference is from the result in the 
  energy buildup phase. However, the difference between CV-A2 and CV-A2D  is only a few percent. From these results, our solution is almost convergent in temporal resolution. 
  According to \cite{2017ApJ...838..113L}, using a higher temporal cadence makes the results of a data-driven simulation closer to the ground-truth data. However, we do not see 
  this in our results. One of possible reason could be due to differences in the methods used to create the simulation.  \cite{2017ApJ...838..113L} directly uses data that are extracted 
  from the ground-truth data where-as our method drives the coronal magnetic field by electric fields given near the solar surface that are derived from the bottom magnetic fields of the 
  ground-truth data. It is therefore likely that these differing methods are the reason for the differing effect to adjusting the temporal resolution of the simulation.
  
  Furthermore, it is important to consider a ratio of the sampling time ($dt_s$) to the dynamical time ($dt_d$) proposed by  \cite{2017ApJ...838..113L}  and the ratio should be less than 1. 
  The dynamical time is defined by  $dt_t$=$\Delta / V_p$ where $\Delta$ is a length of the grid interval and $V_p$ is velocity of the photosphere. In this study, since $\Delta$=$1/320$ and 
  maximum $V_{p(max)}$ = 0.01, $dt_{t(min)}$ is given as 0.3125. The sampling time $dt_s$ is given as 0.5625 (low time resolution cases) and 0.225 (high time resolution cases,), therefore 
  the ratio corresponds to 1.8 and 0.72, respectively. The ratio in the low time resolution cases (other than TW-A2D and CV-A2D) is over 1. However, in this work, the bottom-boundary 
  parameters driving the system are from a model and thoroughly given in form of functions of time and position. From this reason, there is little uncertainty in the boundary data (sampling data) 
  and there is not much difference in energies between the low time resolution cases and high time resolution cases.
  
  \subsection{Data-driven Simulation through the Energy Accumulation Process to the Release Process}
\begin{figure}
  \epsscale{.8}
  \plotone{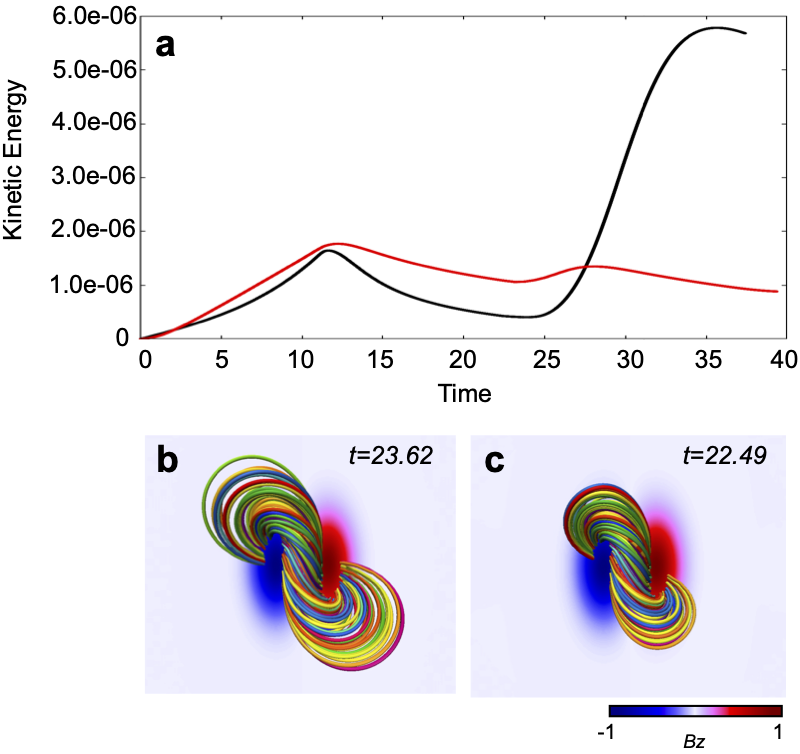}
  \caption{ {\bf a}      The temporal evolution of the kinetic energy from $t=0$ to the last time of the simulation for the ground-truth data 
                                shown in black and the data-driven simulation shown in red, respectively.  This data-driven simulation was conducted with
                                the same manner in TW-A2 and CV-A2.          
                 {\bf b}     Three-dimensional magnetic field lines in pre-erupting phase at $t$=23.26 for the ground-truth data.  
                 {\bf c}      Three-dimensional magnetic field lines in pre-erupting phase for the data-driven simulation. }
  \label{f21}
  \end{figure}

   We run the data-driven simulation throughout the entire process from the energy accumulation process to the release process. The result is shown in Fig. \ref{f21}(a) in which the 
   temporal evolution of the ground-truth data and the data-driven simulation is plotted in black and red, respectively. We found that the data-driven simulation cannot cause the eruption that is 
   different from the results discussed above. One of possible reasons is the pre-erupting magnetic field. Figures \ref{f21}(b) and (c) show the pre-erupting magnetic fields of the ground-truth data 
   and produced by the data-driven simulation, respectively. The size of the sigmoidal structure is different between them. Fig. \ref{f10}(b) shows the magnetic flux of the pre-erupting magnetic 
   fields that depends on the twist value before the relaxation process. From this result, the maximum twist value of the ground-truth data is in a rage from 0.6 to 0.7. On the other hand, the highly 
   twisted field lines buildup in the data-driven simulation are concentrated approximately up to $T_w$=0.6. This may be very small to cause the eruption. For instance, 
    \cite{2018NatCo...9..174I} suggested that to cause the solar flares associated with the eruptions requires the twisted field lines with twist values from 0.5 to 1.0. Therefore, the 
    ground-truth data does not have much of twist and  it would be difficult for the magnetic field produced by the data-driven simulation to cause the eruption. This result indicates that a little difference 
    in twist may affect the eruption. Furthermore, this is a reminder of that the 3D solution is not uniquely determined even using same boundary condition. 

   Although the problem may be caused by the ground-truth data that has weakly twisted field lines before the eruption in this study, we should avoid the problem of weakening of the 
   twist of the magnetic field lines produced in the data-driven simulation when using the photospheric magnetic field. The NLFFF would be useful as the initial condition of the data-driven simulation 
   before the eruption. Because, in some cases, the NLFFF before the eruption reproduces more highly twisted field lines than those shown in Fig. \ref{f21}(c) (\citealt{2016PEPS....3...19I}). 
   However more discussion is needed on the issue of whether to use the magnetic field produced in the data-driven simulation or the NLFFF as the initial condition. This would be addressed in 
   future work.

  \section{Summary}
  The three-dimensional coronal magnetic field plays an essential role in producing solar phenomena such as solar flares, solar jets, and CMEs. However, the 3D coronal magnetic field 
  is not fully understood as only observations of the photospheric magnetic field can currently be made. Data-driven simulations (\citealt{2020ApJ...890..103T}) are a powerful tool to detect 
  not only the 3D magnetic structure but also chart its evolution. However an important question about their reliability still remain. Therefore, in this study, we developed a new data-driven 
  MHD code based on \cite{2018ApJ...855...11H}  which is designed with a zero-beta approximation and a central differential method, and confirmed its performance. In order to check the 
  performance of the developed data-driven MHD simulation code, we carried out a benchmark test using ground-truth data mede from a time evolving 3D magnetic field produced in the 
  MHD simulation (\citealt{2000ApJ...529L..49A} and \citealt{2003ApJ...585.1073A}). This covers the typical subsequent evolution from the energy buildup stage to the erupting stage, as 
  seen in a flare productive active region (\citealt{2019LRSP...16....3T}).  \cite{2017ApJ...838..113L} and \cite{2020ApJ...890..103T} tested the accuracy of their data-driven 
  simulations using the ground-truth data that was made from flux emergence simulations. In this study, our ground-truth data includes the erupting process of the MFR. Consequently, 
  we confirmed that our data-driven simulation can reproduce the erupting processes of the magnetic field well. 
    
  We found that if there are no any treatments for the vertical velocity ($v_z$) at $z=\Delta z$, the temporal evolution deviates strongly from the ground-truth data. 
  This is inferred from the discrepancy due to a derivative in the z derivation of the electric field which includes MHD electric field ($-\vec{v}\times\vec{B}$) and non-MHD electric field 
  ($E^{(3)}$).  In this study, we found that simple linear interpolation or rather $v_z$ set to zero better suppresses the deviation from the original MHD solution. We found that these 
  deviations are larger when the magnetic field evolves in quasi static energy buildup process toward the eruption. After the eruption, we found that the vertical velocity returns 
  towards the values seen in the ground truth data. There are no remarkable differences between the magnetic field evolutions, which are obtained from the data-driven simulations, 
  in the energy buildup phase even though the vertical velocity at $z=\Delta z$  in each case is not the same way. However, from Fig.\ref{f11}b, the initiation of the MFR eruption is later than 
  other cases obtained from the data-driven simulation if the vertical velocity does not behave correctly. This result suggests that the down-flow, which is unlikely in the ground-truth data, 
  inhibits reconnection at the photosphere and hence the formation of  the MFR (see Fig.\ref{f13}a). Therefore, we require the treatment of the vertical velocity at $z=\Delta z$. 
  
  There are several choices for the discretization methods applied at $z=\Delta z$ which is practical boundary condition of the method proposed in \cite{2018ApJ...855...11H}. The critical 
  issue is the derivative in the z direction between the electric fields obtained from the MHD and non-MHD processes. This is because, in general, this derivative is not allowed and that 
  contradiction causes the numerical instability. We suggested that the stability strongly depends on the numerical scheme used. In addition, we here note that the vertical size of the 
  simulation grid ($\Delta z$) can be arbitrarily determined in the simulation setup; therefore, changing $\Delta z$ can be a remedy to reduce the differences in the simulation results 
  of type A and B. We will investigate this point further in the future.
  
    In this study, we found that our data-driven simulation could capture an overview of the evolution of the 3D coronal magnetic field which mimics the evolution as seen in flare productive 
  active region. As a next step, we will apply the observed photospheric magnetic field to our data-driven simulation code. In general, there is no guarantee that the success of this 
  simulation will result in the success of this future simulation. For example, in this study, we assigned enough spatial resolution to the PIL where the converging motion 
  drives the eruption. Although it is important to correctly capture the photospheric motion near the PIL for the eruption, some observational data do not provide that.  Recently, large 
  ground-telescopes such as the GST at BBSO and the Daniel K. Inouye Solar Telescope (\citealt{2021SoPh..296...70R}) provide high resolution 
  data, which will be helpful in resolving this issue. To use these high resolution data, we will need further optimization and implement further technical upgrades in our code. Additionally, 
  in this study, we did not take into account $v_z$ at the bottom surface in the ground-truth data as in \cite{2017ApJ...838..113L} and \citealt{2020ApJ...890..103T} , so we have not estimated 
  how much this velocity effects the data-driven simulation. However, we believe that our data-driven simulation has potential to reproduce the evolution of the magnetic field from the energy 
  buildup stage to its erupting stage even using  observed magnetic field based on its performance in this study.

    \acknowledgments
    We are grateful to a referee for constructive comments and Dr.\ Magnus Woods for reading this paper. This work was supported by the National Science Foundation  under grant 
    AGS-1954737,  AGS-2145253 and National Aeronautics and Space Administration under grant 80NSSC21K1671. This work was also supported by the computational  joint research 
    program of the Institute for Space-Earth Environmental Research, Nagoya University. KH is supported by National Aeronautics and Space Administration under grants  80HQTR20T0067. 
    TM is supported by JSPS KAKENHI Grant Numbers JP20K11851 and JP20H00156. The visualization was done using VAPOR (\citealt{2005SPIE.5669..284C}, 
    \citealt{2007NJPh....9..301C}).    
   
    \section*{Appendix}   
    \appendix
    \section{Derivation of the Poisson Equations}
    \subsection{Derivation of the Second Poisson Equation for $\Phi^{(2)}$}  \label{Apex_PE2}  
     First we multiply $\vec{\nabla}_h \cdot$ by equation (\ref{in_eq2}), then 
    \begin{displaymath}
    \vec{\nabla}_h \cdot (\vec{\nabla}\times\vec{E}^{(2)}) = -\frac{\partial}{\partial z}\left( \frac{\partial E_y^{(2)}}{\partial x}+\frac{\partial E_x^{(2)}}{\partial y}\right)=0,
    \end{displaymath}
    should be satisfied at the boundary. If we assume 
    \begin{equation}
    \vec{E}^{(2)} = -\vec{z}\Phi^{(2)},
    \label{E_2}
   \end{equation}
    the above relationship is satisfied.  We substitute Eq.(\ref{E_2}) to Eq.(\ref{in_eq2}), then  we can obtain 
    \begin{displaymath}
    -\vec{\nabla}\times\vec{E}^{(2)} = -\vec{\nabla}\times(-\vec{z}\Phi^{(2)}) = \left(\frac{\partial \Phi^{(2)}}{\partial y}, -\frac{\partial \Phi^{(2)}}{\partial x},0 \right).
   \end{displaymath}
    Therefore, we describe the equation for $\vec{B}_{h:df}$ as follows, 
    \begin{equation}
    \frac{\partial \vec{B}_{h;df}}{\partial t} =  \left(\frac{\partial \Phi^{(2)}}{\partial y}, -\frac{\partial \Phi^{(2)}}{\partial x}, 0\right).
    \end{equation}
    We multiply $\vec{\nabla}\times$ and its z component corresponds to the second Poisson equation (Eq.\ref{poisson_2}), 
    \begin{displaymath}
    \begin{split}
    \vec{\nabla}^2\Phi^{(2)}& = -\vec{z}\cdot\left(\vec{\nabla}\times\frac{\partial\vec{B}_{h:df}}{\partial t}\right) \\
                                         & = - \left\{\frac{\partial}{\partial x} \left(\frac{\partial B_{y:df}}{\partial t}\right) - \frac{\partial}{\partial y}\left(\frac{\partial B_{x:df}}{\partial t}\right)\right\} \\
                                         & = - \left\{\frac{\partial}{\partial x} \left(\frac{\partial B_{y}}{\partial t}\right) - \frac{\partial}{\partial y}\left(\frac{\partial B_{x}}{\partial t}\right)\right\}.
   \end{split}                                       
   \end{displaymath}
    Since the right-handed term can be derived from the observed $B_x$ and $B_y$, we can find $\Phi^{(2)}$ and eventually obtain $\vec{E}^{(2)}$ through the equation(\ref{E_2}). 
    Note that we used
   \begin{displaymath}
   \frac{\partial}{\partial x} \left(\frac{\partial B_{y:cf}}{\partial t}\right) - \frac{\partial}{\partial y}\left(\frac{\partial B_{x:cf}}{\partial t}\right)= 0. 
    \end{displaymath}
      
    \subsection{Derivation of the Third Poisson Equation for $\Phi^{(3)}$} \label{Apex_PE3}
    The solenoidal condition, $\vec{\nabla}\cdot\vec{B}=0$ can be rewritten as follows, 
    \begin{displaymath}
    \vec{\nabla}\cdot\vec{B}_h  + \frac{\partial B_z}{\partial z} = 0, 
    \end{displaymath}
    and we obtain the following relationship by conducting the time derivative of the above equation,
    \begin{equation}
    \begin{split}
    \frac{\partial}{\partial z}\left(\frac{\partial B_z}{\partial t}\right) & = -\vec{\nabla}_h\cdot\left(\frac{\partial \vec{B}_h}{\partial t}\right) \\
                                                                                                    & = -\vec{\nabla}_h\cdot\left(\frac{\partial \vec{B}_{h:cf}}{\partial t}\right) \\
                                                                                                    & = -\vec{\nabla}_h\cdot \left\{-(\vec{\nabla}\times\vec{E}^{(3)})_h \right\} \\
                                                                                                    & = -\frac{\partial}{\partial z}(\vec{\nabla}\times\vec{E}^{(3)})\cdot\vec{z}, 
    \label{eq_e3_a1}                                                                                                 
    \end{split}
    \end{equation}
    where we used  $ \vec{\nabla}\cdot (\vec{\nabla}\times\vec{E}^{(3)}) = 0$. Eq.(\ref{eq_e3_a1}) can be described by taking into account Eq.(\ref{eq_e3_1}), as follows, 
    \begin{displaymath}
    \frac{\left(\frac{\partial B_z^{(3)}}{\partial t}\right)_{z=+\frac{1}{2}\Delta z} - \left(\frac{\partial B_z^{(3)}}{\partial t}\right)_{z=0}}{\frac{1}{2}\Delta z} 
    = -\vec{\nabla}_h\cdot\left(\frac{\partial\vec{B}_{h:cf}}{\partial t}\right).
    \end{displaymath}
    Since $B_z^{(3)}$ should be zero at the photosphere to avoid overlapping with $B_z^{(1)}$,  
    \begin{equation}
    \left(\frac{\partial B_z^{(3)}}{\partial t}\right)_{z=+\frac{1}{2}\Delta z} = - \frac{1}{2}\Delta z \vec{\nabla}_h\cdot\left(\frac{\partial \vec{B}_{h:cf}}{\partial t}\right).
    \label{eq_e3_3}
    \end{equation}
    When we make same assumption as when we obtained $\vec{E}^{(1)}$, 
    \begin{equation}
    \vec{E}^{(3)}|_{z=+\frac{1}{2}\Delta z} = \vec{z} \times \vec{\nabla}_h\Phi^{(3)}=\left(-\frac{\partial \Phi^{(3)}}{\partial y}, \frac{\partial \Phi^{(3)}}{\partial x}, 0 \right), 
    \label{eq_e3_4}
    \end{equation}
     the equation
     \begin{equation}
      \vec{z}\cdot(\vec{\nabla}\times\vec{E}^{(3)})  = \left(\frac{\partial^2 \Phi^{(3)}}{\partial x^2} + \frac{\partial^2 \Phi^{(3)}}{\partial y^2}\right) 
                                                                             = -\frac{\partial B_z^{(3)}}{\partial t},
     \label{eq_e3_5}
     \end{equation}
     is obtained from equation(\ref{eq_e3_1}). Eventually we can obtain the third poisson equation, 
     \begin{displaymath}
     \vec{\nabla}_h^2 \Phi^{(3)} = \frac{1}{2}\Delta z \vec{\nabla}_h\cdot\left(\frac{\partial \vec{B}_{h:cf}}{\partial t}\right).
     \end{displaymath}

    \section{Performances of the Poisson Equations}
    We show the performances of the Poisson equations (\ref{poisson_1}), (\ref{poisson_2}), and (\ref{poisson_3}) by evaluating the following values,
    \begin{displaymath}
    L_1 = \int\left| \vec{\nabla}^2 \Phi^{(1)} - \frac{\partial B_z}{\partial t}\right|dV,  
    \end{displaymath}
    \begin{equation}
    L_2 = \int\left|\vec{\nabla}^2\Phi^{(2)} -\vec{z}\cdot\left(\vec{\nabla}\times\frac{\partial\vec{B}_{h:df}}{\partial t}\right)\right|dV,
    \end{equation}
    \begin{displaymath}
    L_3 = \int\left|\vec{\nabla}_h^2 \Phi^{(3)} - \frac{1}{2}\Delta z \vec{\nabla}_h\cdot\left(\frac{\partial \vec{B}_{h:cf}}{\partial t}\right)\right|dV,
    \end{displaymath}
    and
     \begin{displaymath}
    ER_1 = \left| \vec{\nabla}^2 \Phi^{(1)}\right|_{max} - \left|\frac{\partial B_z}{\partial t}\right|_{max},  
    \end{displaymath}
    \begin{equation}
    ER_2 = \left|\vec{\nabla}^2\Phi^{(2)}\right|_{max} - \left|\vec{z}\cdot\left(\vec{\nabla}\times\frac{\partial\vec{B}_{h:df}}{\partial t}\right)\right|_{max},
    \end{equation}
    \begin{displaymath}
    ER_3 = \left|\vec{\nabla}_h^2 \Phi^{(3)}\right|_{max} - \left|\frac{1}{2}\Delta z \vec{\nabla}_h\cdot\left(\frac{\partial \vec{B}_{h:cf}}{\partial t}\right)\right|_{max}.
    \end{displaymath}
    When  the solutions of the each Poisson equation are correctly obtained, these values completely drop to zero. The Poisson equation is solved numerically by a simple Gauss-Seidel 
    method based on the second order finite difference method and the initial  $\Phi$ is given as zero. Figures \ref{fA_1}a and b show the iteration profile of $L_n$ and $ER_n$,  
    respectively, where n=1, 2, 3.  We find that the each value dramatically decreases during the each iteration and eventually saturates at a very low value compared to the initial value. 
    Therefore, the solutions would be obtained with good accuracy. 
    
    \section{Reproducibility of the Photospheric Magnetic Field}
     We check how much  reproducibility of  the photospheric magnetic field from the electric field through the equations (\ref{eq_bottom}) has. Following \cite{2018ApJ...855...11H},  
   the bottom magnetic field should be reproduced perfectly through the data-driven simulation but it depends on the accuracy of the Poisson solver. Following them, we also  make scatter 
    plots for the photospheric magnetic field of the ground-truth data vs. the magnetic field reproduced via the data-driven simulation using the electric fields. Figure \ref{fA_2} shows the 
    scatter plots for $B_x$ and $B_y$ in the twisting phase from t=0 $\sim$ t=11.25. Note that the $B_z$ component is not plotted because this does not change during this time period. 
    These plots are almost along a function,  $y=x$, {\it i.e.}, the magnetic field is reproduced well from the data-driven simulation. Figure \ref{fA_3} shows these scatter plots in the erupting 
    phase, which are the same format as in Fig. \ref{fA_2} except $B_z$ is plotted. Although there's a little dispersions compared to previous cases (however the correlation coefficient 
    is over 0.99), the scatter plots are mostly along a line of $y=x$. So, throughout the simulation, the boundary magnetic field is reproduced well from the data-driven simulation based 
    on the given electric fields.

\clearpage

  \clearpage

  \clearpage
  
  \clearpage

  \clearpage

  \clearpage

  \clearpage

  \clearpage
  
  \clearpage
  
  \clearpage

  \clearpage

  \clearpage
  
  \clearpage

  \clearpage

  \clearpage

  \clearpage

  \clearpage

  \clearpage  

  \clearpage  

  \clearpage  
  
  \clearpage  
  
  \clearpage

    \appendix
    \renewcommand{\thefigure}{A\arabic{figure}}
    \setcounter{figure}{0}
  \begin{figure}
  \epsscale{1.}
  \plotone{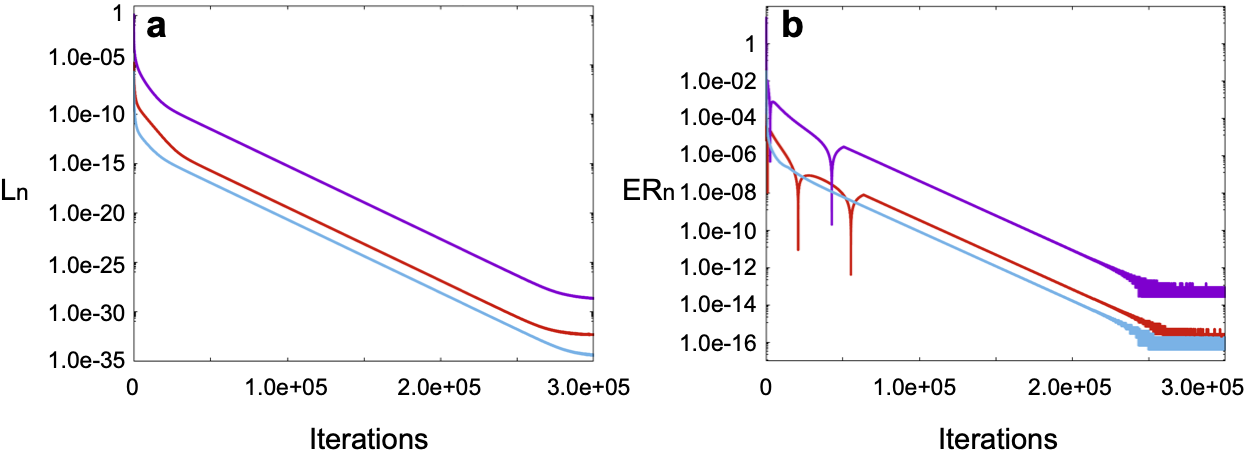}
  \caption{
                (a)-(b) Evolutions of $L_n$ and $ER_{n}$, where n=1,2,3, during the iterations of each Poisson equation. Each 
                color, red, purple, and blue represent n=1, n=2, and n=3, respectively.
                }
  \label{fA_1}
  \end{figure}
  \clearpage    
  
  \begin{figure}
  \epsscale{.9}
  \plotone{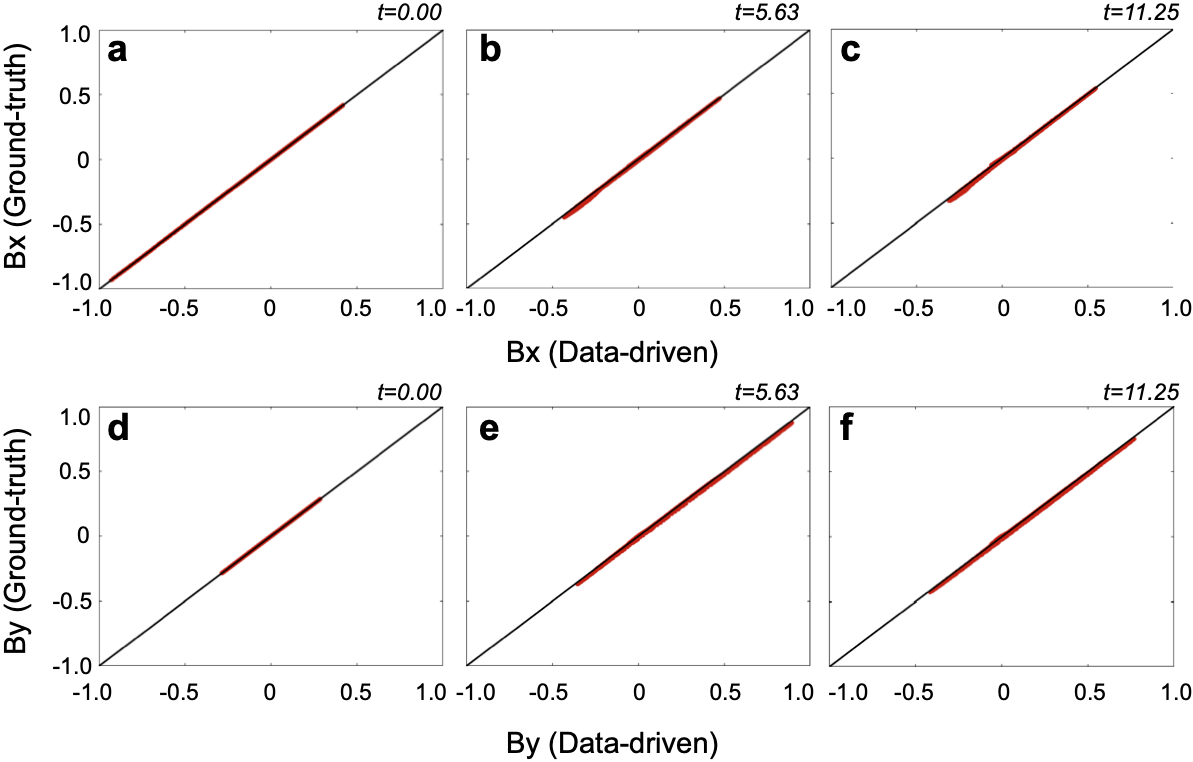}
  \caption{ The scatter plot of the bottom magnetic fields by the ground-truth data vs. reproduced through the data-driven 
                simulation. The upper and lower panels correspond to the distributions of $B_x$ 
                 and $B_y$, respectively,  at each time under the twisting motion.  Note that the scatter plots on $B_z$  are excluded because 
                 it does not changed during the evolution.     
                               }
  \label{fA_2}
  \end{figure}
  \clearpage

  \begin{figure}
  \epsscale{1.}
  \plotone{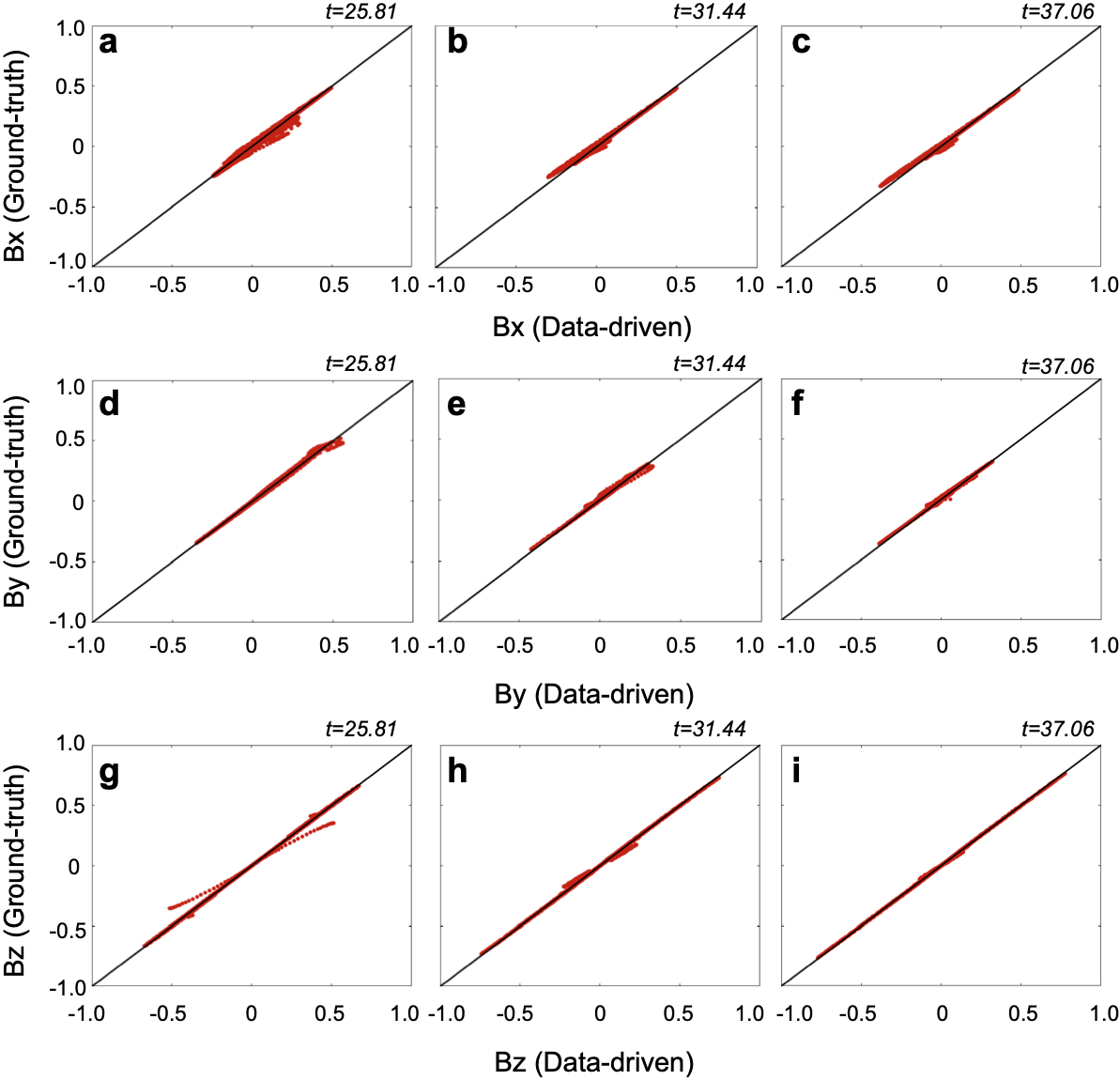}
  \caption{
                The scatter plot of the bottom magnetic fields from the ground-truth data vs. that reproduced through the data-driven simulation during
                the erupting stage, which are presented in the same format as in Fig.\ref{fA_2}. The upper, middle and lower panels are the results 
                regarding to $B_x$, $B_y$, and $B_z$ components, respectively. Although these look a little scattered compared to Fig. \ref{fA_2},  
                the correlation coefficients are over 0.99.}
  \label{fA_3}
  \end{figure}
  \clearpage

    \begin{table}
    \begin{center}
    \caption{All cases of the data-driven MHD simulation undertaken in this study are shown in the 1st column. The type of differential method and a way of 
    updating for $v_z$ at k=1 are listed in the 2nd and 3rd columns. Interpolation means that $v_z$ at k=1 is derived from a linear interpolation with 
    values at k=0 and k=2, {\it i.e.}, $v_z|_{k=1} = 0.5v_z|_{k=2}$. Because $v_z|_{k=0}$=0 is set in this study. Case, TW-A2D and CV-A2D are 
    given with 2.5 times the temporal resolution of TW-A2 and CV-A2, respectively.  \label{tbl-1}}
    \begin{tabular}{crr}
    \tableline\tableline
    Case & Differential Method & $v_z$ at k=1 \\
   \tableline
   TW-A0      & A &  Equation of motion     \\
   TW-A1      & A &  zero                            \\
   TW-A2      & A & interpolation                \\
   TW-A2D    & A & interpolation                \\
   TW-B        & B & interpolation                \\
   CV-A0       & A & Equation of motion      \\
   CV-A1       & A & zero                             \\
   CV-A2       & A & Interpolation                \\
   CV-A2D    & A & Interpolation                \\
   CV-B        &B&  Interpolation                    \\
   \tableline
   \end{tabular}
   \end{center}
   \end{table}

\end{document}